\documentclass[10pt,onecolumn,english,aps,prd,superscriptaddress,nofootinbib,preprintnumbers,floatfix,longbibliography]{revtex4-2}

\usepackage[utf8]{inputenc}
\usepackage[english]{babel}
\usepackage[T1]{fontenc}
\usepackage{graphicx}
\usepackage{amsmath,amssymb,amsfonts,bm}
\usepackage{dcolumn}
\usepackage{makecell}
\usepackage{xcolor}
\usepackage{multirow}
\usepackage{placeins}
\usepackage{mathrsfs}
\usepackage[colorlinks=true,linkcolor=blue,citecolor=blue,urlcolor=blue]{hyperref}
\usepackage{orcidlink}

\begin{document}

\title{Blandford--Znajek Scaling in a Power-Law Rotating Kalb--Ramond Geometry: Magnetic-Flux Systematics and Bayesian Identifiability}

\author{Sardor~Murodov\orcidlink{0000-0003-2360-4475}}
\email{s.murodov@newuu.uz}
\affiliation{New Uzbekistan University, Movarounnahr Street 1, Tashkent 100000, Uzbekistan}
\affiliation{Tashkent State Technical University, Tashkent 100095, Uzbekistan}

\author{Olimjon Kholturayev\orcidlink{0009-0004-3758-9610}} 
\email{o.xoltorayev@newuu.uz}
\affiliation{New Uzbekistan University, Movarounnahr Street 1, Tashkent 100000, Uzbekistan}

\author{Bekzod~Rahmatov\orcidlink{0009-0001-0394-650X}}
\email{rahmatovbekzod@samdu.uz}
\affiliation{University of Tashkent for Applied Sciences, Str. Gavhar 1, Tashkent 100149, Uzbekistan}

\author{Javlon~Rayimbaev\orcidlink{0000-0001-9293-1838}}
\email{javlon@astrin.uz}
\affiliation{Institute of Theoretical Physics, National University of Uzbekistan, Tashkent 100174, Uzbekistan}
\affiliation{Kimyo International University in Tashkent, Shota Rustaveli street 156, Tashkent 100121, Uzbekistan}

\author{Islom Egamberdiev\orcidlink{0009-0001-6156-9271}}
\email{egamberdiyev.islom@samdaqu.edu.uz}
\affiliation{Samarkand State Technical University named after Mirzo Ulugbek, Lolazor street 70, Samarqand 140143, Uzbekistan}

\author{Shavkat Karshiboev\orcidlink{0009-0006-4847-5947}}
\email{shavkat.qarshiboyev.89@bk.ru}
\affiliation{Samarqand State Pedagogical Institute, Spitamen Shokh Street 166, Samarkand 140100, Uzbekistan}

\date{\today}

\begin{abstract}
Relativistic jets from spinning black holes offer a possible strong-field probe of gravity through the Blandford--Znajek mechanism. We study the leading jet-power scaling in the four-dimensional power-law rotating Kalb--Ramond geometry introduced by Kumar, Ghosh, and Wang. The metric is used here as a stationary background, without assuming that it constitutes a newly established exact rotating solution. We first examine the deformation at the metric level. For $s=2$ it is absorbed completely by a mass redefinition, whereas for $s>2$ the correction decays more slowly than the usual mass term. Our main benchmark is therefore the nondegenerate $s=3/2$ case, whose correction falls faster than the Kerr mass term; $s=3$ is kept as a secondary comparison. We evaluate the BZ scaling under three magnetic assumptions: fixed total horizon flux, fixed local normal field with the proper horizon area, and a reduced radius-based flux proxy. The resulting trends differ appreciably, showing that the magnetic prescription is itself a leading systematic. For GRO~J1655--40 and GRS~1915+105, the marginalized deformation posterior remains close to the horizon-conditioned effective prior for both a uniform prior and a truncated-Gaussian alternative. The jet-only profile likelihood is also nearly flat over the allowed deformation range, with the same qualitative behavior in the $s=3$ test. Thus, within the present setup, the jet-power proxies do not independently determine the Kalb--Ramond deformation. A stronger inference will require better control of the rotating background, source-dependent magnetic flux, non-Kerr spin estimates, and a larger sample.
\end{abstract}

\maketitle

\section{Introduction}

Relativistic jets from accreting black holes rank among the most energetic collimated outflows in the Universe. In black-hole X-ray binaries, the empirical connection between accretion state and jet production is well established \cite{Fender2001,Fender2004}. Galactic microquasars have also provided resolved examples of relativistic and apparently superluminal ejecta, most notably GRS~1915+105 and GRO~J1655--40 \cite{MirabelRodriguez1994,HjellmingRupen1995}. More recent multiwavelength campaigns continue to refine the phenomenology of transient ejections and disc--jet coupling \cite{Zdziarski2022,Carotenuto2022}. Horizon-scale observations of M87* and Sgr~A* by the Event Horizon Telescope have likewise highlighted the need to model accretion, magnetic fields, and spacetime geometry together \cite{EHT2019M87I,EHT2019M87V,EHT2022SgrAI}. Jets add a complementary probe of the near-horizon region of rotating black holes.

A standard framework for powering such outflows is the Blandford--Znajek (BZ) mechanism, where magnetic fields threading a rotating black hole extract rotational energy electromagnetically \cite{Blandford1977,McKinney2005,Garofalo2009}. Its spin dependence and horizon-scale energetics have been tested and refined analytically and numerically \cite{Tchekhovskoy2010,NarayanPennaSadowski2013,Penna2014,Penna2015}. Force-free and GRFFE treatments establish the roles of horizon regularity, field-line rotation, and magnetic-field geometry \cite{Komissarov2001,McKinney2006,McKinneyNarayan2007,KomissarovMcKinney2007}, while higher-order and dynamical analyses clarify the structure of the magnetosphere and the physical origin of the outgoing Poynting flux \cite{PanYu2015,Noda2020,KoideImamura2019,LyutikovMcKinney2011}. Related analytical developments also emphasize the global structure and regularity of force-free solutions \cite{Camilloni2022}. In the standard Kerr spacetime, the leading BZ power is controlled by the horizon magnetic flux and horizon angular velocity. In the high-magnetic-flux regime, GRMHD simulations show that magnetically arrested accretion can produce very efficient jets \cite{Tchekhovskoy2011,McKinneyTchekhovskoyBlandford2012,WhiteStoneQuataert2019}. Reconnection, radiative effects, plasma supply, and magnetic-field topology can further regulate the emerging outflow \cite{TchekhovskoyMcKinneyDexter2014,AvaraMcKinneyReynolds2016,Parfrey2019,Singh2018}. Magnetocentrifugal and collapsar models provide complementary jet-launching settings \cite{BlandfordPayne1982,BarkovKomissarov2008}. A change in the horizon or frame-dragging structure can therefore feed directly into the electromagnetic extraction rate.

The BZ mechanism has been extended beyond Kerr to several non-Kerr and modified-gravity backgrounds. Early and recent analyses show that deviations from the Kerr geometry can modify the BZ power and force-free magnetosphere \cite{Pei2016,Dong2022,Camilloni2024}. Quantum- or deformation-inspired black-hole models provide additional examples in which the jet efficiency departs from the Kerr prediction \cite{Dihingia2024,Banerjee2021}. Jet energetics can therefore be used alongside black-hole shadows, accretion-flow images, and gravitational-wave observables when testing the spacetime geometry.

Kalb--Ramond (KR) gravity provides a theoretically motivated framework in which an antisymmetric tensor field can induce Lorentz-violating corrections to the spacetime geometry. The KR two-form originates in the classic string-inspired construction \cite{KalbRamond1974}, while spontaneous Lorentz breaking by tensor vacuum expectation values and the corresponding Standard-Model Extension framework were developed in Refs.~\cite{KosteleckySamuel1989,ColladayKostelecky1997,ColladayKostelecky1998,Kostelecky2004}. Phenomenological consequences and constraints on Lorentz-violating tensor backgrounds have been investigated in several complementary settings \cite{Altschul2010,Lambiase2005,Das2018,Mavromatos2014}, including KR-torsion applications \cite{deCesare2015}.

Static and charged KR black-hole solutions provide the basis for much of the recent strong-field phenomenology \cite{Lessa2020,Liu2024Static,Duan2024,LiuWuWei2025}. Rotating KR geometries and their optical signatures have been studied in both power-law and slowly rotating constructions \cite{Kumar2020,Liu2025}. Recent extensions include particle dynamics, thermodynamics, evaporation, and neutrino propagation in KR backgrounds \cite{Araujo2025Particle,AraujoFilho2025ParticleMotionKR,Shi2025NeutrinoKR,AraujoCQG2025}. The optical sector has also been developed for charged, noncommutative, and nonlinear-electrodynamic KR spacetimes through lensing, time-delay, and related propagation observables \cite{Araujo2025Noncomm,AraujoFilho2025AntisymmetricTensorLensing,Pereira2026ChargedKRLensing,AraujoFilho2026OpticalNonCommutativeKR,Mangut2025}, while perturbative and accretion properties have been considered in Refs.~\cite{Gu2026,Zulqarnain2026}.

Complementary strong-field applications include circular motion and QPO phenomenology in KR backgrounds \cite{Murodov2024KR,Murodov2025ChargedKR}, PFDM and lensing environments \cite{Murodov2025PFDMKR,Murodov2025LensingKR}, and related non-Kerr orbital studies \cite{Murodov2023Universe,Murodov2024CJP,Murodov2026KS}. Observational signatures and polarized images of rotating charged KR black holes have also been considered using thin accretion-disk models \cite{Yang2026}. These results establish that KR deformations can influence the horizon structure as well as particle, photon, and electromagnetic observables near compact objects.

Energy extraction in Kalb--Ramond gravity has also recently been explored. In particular, Yao et al. studied the Comisso--Asenjo magnetic-reconnection mechanism for a distinct rotating charged KR black-hole family and quantified how its Lorentz-violating parameter changes the allowed extraction region, power, and efficiency \cite{Yao2026}. That mechanism is physically complementary to the force-free Blandford--Znajek process considered here. Our purpose is not to claim a new exact rotating KR solution or a precision jet-based bound on the underlying gravity theory. Instead, we ask whether the adopted power-law rotating geometry contains an independently identifiable non-Kerr parameter and, if it does, whether a leading BZ analysis can distinguish it from assumptions about the horizon magnetic flux and the source priors.

We organize the analysis around three checks. At the metric level, the $s=2$ member reduces to Kerr after a mass redefinition, whereas the large-radius correction scales as $r^{-2/s}$ for positive $s$; hence $s>2$ decays more slowly than the usual $1/r$ mass term. We use $s=3/2$, with its $r^{-4/3}$ correction, as the main nondegenerate benchmark and keep the literature-used $s=3$ case as a secondary test. We then compare fixed total horizon flux, fixed local normal field evaluated with the proper horizon area, and a reduced radius-based flux proxy. Finally, using GRO~J1655--40 and GRS~1915+105, we compare the deformation posterior with the horizon-conditioned effective prior and with a profile likelihood that does not depend on the deformation prior.

Accordingly, the calculation should be read as a horizon-based BZ scaling test on an adopted stationary geometry. We do not solve a KR-specific Grad--Shafranov, GRFFE, or GRMHD problem, nor do we establish that the rotating metric together with a rotating two-form satisfies the complete Einstein--KR equations. The numerical and Bayesian results are interpreted within these limits.

The paper is organized as follows. Section~\ref{sec:KR_spacetime} introduces the adopted rotating KR geometry, including the metric degeneracy and asymptotic caveats. Section~\ref{sec:force_free_magnetosphere} summarizes the stationary force-free setup, and Sec.~\ref{sec:BZ_power_KR} develops the leading BZ scaling and the magnetic-flux prescriptions. Section~\ref{sec:numerical_analysis} compares their numerical consequences. Section~\ref{sec:MCMC_constraints} presents the Bayesian consistency and identifiability analysis, including the horizon-conditioned effective-prior and profile-likelihood tests. Astrophysical implications are discussed in Sec.~\ref{sec:astrophysical_implications}, followed by the conclusions in Sec.~\ref{sec:conclusions}.

\section{Rotating Kalb--Ramond black hole spacetime}
\label{sec:KR_spacetime}

We begin by specifying the gravitational setup and the rotating Kalb--Ramond geometry used below. The Kalb--Ramond (KR) field is described by an antisymmetric rank-two tensor field \(B_{\mu\nu}\), originating in string-inspired constructions \cite{KalbRamond1974} and capable of inducing spontaneous Lorentz symmetry breaking through a nonzero vacuum expectation value \cite{KosteleckySamuel1989,Kostelecky2004,Altschul2010}. Explicit KR black-hole solutions demonstrate how such a background modifies the spacetime geometry \cite{Lessa2020,Liu2024Static}. Since the Blandford--Znajek mechanism is sensitive to the horizon and frame-dragging structure, these geometric corrections can affect the jet power.

We do not rederive the rotating metric from the full Einstein--Kalb--Ramond field equations. Instead, we adopt the four-dimensional Kerr-like power-law geometry of Kumar, Ghosh, and Wang \cite{Kumar2020} and examine its implications for electromagnetic energy extraction. The source construction starts from a static power-law hairy KR black hole obtained from the modified field equations and then introduces a stationary, axisymmetric counterpart. Since we do not independently substitute the rotating metric and a rotating two-form configuration back into the complete Einstein--KR equations, throughout this paper we refer to Eq.~\eqref{eq:rotating_KR_metric} as an \emph{adopted rotating KR geometry} and refrain from making a stronger field-equation-exactness claim. A distinct slowly rotating KR solution has been derived consistently as a first-order-in-spin expansion \cite{Liu2025}; it is cited here for comparison but is not the background used below. The two constructions need not describe the same rotating branch of the theory.

\subsection{Action and field content}

We first summarize the effective Einstein--Kalb--Ramond action that motivates the static seed and fixes the parameter interpretation of the power-law family \cite{Lessa2020,Kumar2020,LiuWuWei2025}. This action is included to specify the theoretical origin of the background; it should not be read as a new derivation of the rotating metric used later,
\begin{equation}
S=\int d^4x\sqrt{-g}
\left[
\frac{R}{2\kappa}
-\frac{1}{12}H_{\lambda\mu\nu}H^{\lambda\mu\nu}
-V(X)
+\mathcal{L}_{\rm int}
\right],
\label{eq:KR_action}
\end{equation}
where \(\kappa=8\pi G\), \(R\) is the Ricci scalar, \(g\) is the determinant of the metric tensor \(g_{\mu\nu}\), and \(H_{\lambda\mu\nu}\) is the field strength of the KR tensor field. It is defined by
\begin{equation}
H_{\lambda\mu\nu}
=
\partial_{\lambda}B_{\mu\nu}
+
\partial_{\mu}B_{\nu\lambda}
+
\partial_{\nu}B_{\lambda\mu}.
\label{eq:H_field}
\end{equation}

The potential \(V(X)\) is responsible for triggering spontaneous Lorentz symmetry breaking. The argument of the potential can be written as
\begin{equation}
X=B_{\mu\nu}B^{\mu\nu}\pm b^2 ,
\label{eq:potential_argument}
\end{equation}
where \(b^2\) is a constant associated with the vacuum value of the antisymmetric field. At the vacuum, the KR field satisfies
\begin{equation}
\langle B_{\mu\nu}\rangle=b_{\mu\nu},
\qquad
b_{\mu\nu}b^{\mu\nu}=\mp b^2,
\label{eq:KR_vev}
\end{equation}
and the potential obeys
\begin{equation}
V=0,
\qquad
V'=0.
\label{eq:vacuum_conditions}
\end{equation}
Thus, the vacuum tensor \(b_{\mu\nu}\) selects preferred directions in spacetime and leads to Lorentz-violating corrections in the gravitational sector.

The interaction term \(\mathcal{L}_{\rm int}\) represents possible nonminimal couplings between the KR field and curvature. A commonly used effective form is
\begin{equation}
\mathcal{L}_{\rm int}
=
\frac{\xi_2}{2\kappa}
B^{\lambda\nu}B^{\mu}{}_{\nu}R_{\lambda\mu}
+
\frac{\xi_3}{2\kappa}
B_{\mu\nu}B^{\mu\nu}R ,
\label{eq:nonminimal_coupling}
\end{equation}
where \(\xi_2\) and \(\xi_3\) are coupling constants. These terms encode the effect of the antisymmetric tensor background on the spacetime geometry.

Variation of the action with respect to the metric gives modified gravitational field equations of the general form
\begin{equation}
G_{\mu\nu}
=
\kappa T_{\mu\nu}^{\rm KR}
+
\mathcal{C}_{\mu\nu}^{\rm KR},
\label{eq:modified_field_equations}
\end{equation}
where \(T_{\mu\nu}^{\rm KR}\) is the stress-energy tensor of the KR field and \(\mathcal{C}_{\mu\nu}^{\rm KR}\) denotes the correction terms coming from the nonminimal curvature couplings. In the limit where the KR field contribution vanishes, Eq.~\eqref{eq:modified_field_equations} reduces to the standard Einstein field equations. In the power-law static solution underlying Ref.~\cite{Kumar2020}, the source parameter is related to the nonminimal coupling through
\begin{equation}
s=|b^2|\,\xi_2,
\label{eq:s_coupling_mapping}
\end{equation}
so that \(s\) is itself a dimensionless Lorentz-violating combination. This relation is important for interpreting the numerical slice adopted below.

\subsection{Metric and horizon structure}

The metric used in this work is the four-dimensional power-law rotating KR geometry proposed by Kumar, Ghosh, and Wang \cite{Kumar2020}. Related static, charged, and slowly rotating KR black-hole solutions are discussed in Refs.~\cite{Lessa2020,Duan2024,Liu2024Static,Liu2025}, while their orbital and radiative phenomenology has been explored in Refs.~\cite{Murodov2024KR,Murodov2025ChargedKR}. We keep the source notation and denote the independent power-law hair amplitude by $\Gamma$. The Kerr-like stationary and axisymmetric line element in Boyer--Lindquist coordinates is
\begin{equation}
ds^2
=
g_{tt}dt^2
+
2g_{t\phi}dtd\phi
+
g_{rr}dr^2
+
g_{\theta\theta}d\theta^2
+
g_{\phi\phi}d\phi^2 .
\label{eq:general_axisymmetric_metric}
\end{equation}
A useful form is
\begin{align}
ds^2
=&
-\left(1-\frac{2Mr-\Gamma r^{n}}{\Sigma}\right)dt^2
-\frac{2a(2Mr-\Gamma r^{n})\sin^2\theta}{\Sigma}dtd\phi
+\frac{\Sigma}{\Delta_{\rm KR}}dr^2
+\Sigma d\theta^2
\nonumber\\
&+
\left[
r^2+a^2
+
\frac{a^2(2Mr-\Gamma r^{n})\sin^2\theta}{\Sigma}
\right]
\sin^2\theta d\phi^2 ,
\label{eq:rotating_KR_metric}
\end{align}
where
\begin{equation}
\Sigma=r^2+a^2\cos^2\theta ,
\label{eq:sigma}
\end{equation}
and
\begin{equation}
\Delta_{\rm KR}
=
r^2-2Mr+a^2+\Gamma r^{n}.
\label{eq:delta_KR}
\end{equation}
Here $M$ and $a$ are the mass and rotation scales appearing in the metric, while $\Gamma$ is the power-law hair amplitude of Ref.~\cite{Kumar2020}; it is not the fundamental nonminimal coupling $\xi_2$. The radial exponent is
\begin{equation}
n=\frac{2(s-1)}{s},
\label{eq:n_parameter}
\end{equation}
with $s=|b^2|\xi_2$ as in Eq.~\eqref{eq:s_coupling_mapping}. Dimensional consistency gives $[\Gamma]=L^{2-n}=L^{2/s}$. At fixed $s$, a convenient dimensionless amplitude is therefore
\begin{equation}
\bar{\Gamma}=\Gamma M^{n-2}.
\label{eq:dimensionless_gamma_general}
\end{equation}
The Kerr limit is $\Gamma=0$.

\subsubsection{Degenerate $s=2$ slice and nondegenerate benchmark}
\label{subsubsec:s2_degeneracy}

A metric-level identifiability check is essential before using the hair parameter observationally. For $s=2$, Eq.~\eqref{eq:n_parameter} gives $n=1$, and the two combinations carrying the deformation reduce to
\begin{align}
2Mr-\Gamma r &= 2M_{\rm eff}r,\\
\Delta_{\rm KR} &= r^2-2M_{\rm eff}r+a^2,
\end{align}
where
\begin{equation}
M_{\rm eff}\equiv M-\frac{\Gamma}{2}.
\label{eq:Meff_s2}
\end{equation}
Because every occurrence of $M$ and $\Gamma$ in Eq.~\eqref{eq:rotating_KR_metric} enters through these combinations, the complete $s=2$ geometry is exactly the Kerr metric written with the redefined mass $M_{\rm eff}$. Hence $\Gamma$ is not independently identifiable on this slice: an $n=1$ Bayesian ``constraint'' would amount to constraining a mass reparameterization rather than a genuine non-Kerr degree of freedom.

The large-radius behavior provides a second identifiability criterion. From the static seed and the rotating metric, the leading deformation of $g_{tt}$ scales as $\Gamma/r^{2/s}$. For positive $s$, a conventional mass-dominated asymptotic expansion therefore requires $2/s>1$, or $s<2$. The borderline value $s=2$ is precisely the algebraically degenerate case identified above, whereas $s>2$ decays more slowly than the usual $1/r$ mass term and requires additional care in defining standard asymptotic charges.

We therefore choose the nondegenerate member
\begin{equation}
s=\frac{3}{2},
\qquad
n=\frac{2}{3},
\qquad
[\Gamma]=L^{4/3},
\qquad
\bar{\Gamma}\equiv\frac{\Gamma}{M^{4/3}}
\label{eq:s15_benchmark}
\end{equation}
as the primary benchmark for all numerical and Bayesian results below. The value $s=3/2$ is not singled out by the observations; it is a representative noninteger member of the asymptotically mass-dominated interval $0<s<2$, chosen away from both the Kerr--Newman-like $s=1$ case of Ref.~\cite{Kumar2020} and the exactly Kerr-reparameterizable $s=2$ boundary. Its leading deformation in $g_{tt}$ scales as $\bar{\Gamma}(M/r)^{4/3}$ and therefore falls faster than the Kerr mass term. This choice does not by itself prove the standard ADM-charge construction for the adopted rotating metric, but it avoids the slower-than-$1/r$ falloff present for $s>2$ and leaves $M$ as the leading asymptotic mass scale.

For comparison with the original shadow literature, which used $s=3$ as a representative case \cite{Kumar2020}, we also retain
\begin{equation}
s=3,
\qquad
n=\frac{4}{3},
\qquad
\bar{\Gamma}_{3}\equiv\frac{\Gamma}{M^{2/3}}.
\label{eq:s3_benchmark}
\end{equation}
In that member the deformation behaves as $r^{-2/3}$, slower than the standard mass term. We therefore use it only as a benchmark-specific robustness comparison and do not interpret its normalization scale as a precision ADM-mass measurement. Because $\Gamma$ has different mass dimension for different $s$, $\bar{\Gamma}$ and $\bar{\Gamma}_{3}$ are distinct dimensionless amplitudes and must not be compared as if they represented one universal coupling. The two benchmark calculations use the same dimensionless prior interval solely to test the robustness of the statistical-identifiability conclusion.

The event horizons are determined by the condition
\begin{equation}
g^{rr}=0,
\label{eq:horizon_condition}
\end{equation}
or equivalently
\begin{equation}
\Delta_{\rm KR}(r)=0.
\label{eq:delta_zero}
\end{equation}

Therefore, the horizon radii are obtained from
\begin{equation}
r^2-2Mr+a^2+\Gamma r^{n}=0.
\label{eq:horizon_equation}
\end{equation}

The largest real positive root of Eq.~\eqref{eq:horizon_equation} defines the outer event horizon,
\begin{equation}
r_H=r_+ .
\label{eq:outer_horizon}
\end{equation}

In the Kerr limit \(\Gamma=0\), Eq.~\eqref{eq:horizon_equation} becomes
\begin{equation}
r^2-2Mr+a^2=0,
\label{eq:kerr_horizon_equation}
\end{equation}
and the usual Kerr horizons are recovered:
\begin{equation}
r_{\pm}^{\rm Kerr}
=
M\pm\sqrt{M^2-a^2}.
\label{eq:kerr_horizons}
\end{equation}

This limit is important because it confirms that the adopted KR geometry continuously reduces to the standard Kerr geometry when the Lorentz-violating correction is switched off.

\subsection{Ergosphere and horizon angular velocity}

The ergosphere is defined by the surface where the timelike Killing vector becomes null. This condition is given by
\begin{equation}
g_{tt}=0.
\label{eq:ergosphere_condition}
\end{equation}

Using Eq.~\eqref{eq:rotating_KR_metric}, the static limit surface is determined from
\begin{equation}
1-\frac{2Mr-\Gamma r^{n}}{\Sigma}=0,
\label{eq:static_limit}
\end{equation}
or
\begin{equation}
r^2+a^2\cos^2\theta
=
2Mr-\Gamma r^{n}.
\label{eq:ergosphere_equation}
\end{equation}

The region between the event horizon \(r_H\) and the static limit surface is the ergoregion. Inside this region, no observer can remain static with respect to infinity because of the strong frame-dragging effect. This property is essential for rotational energy extraction mechanisms, including the Penrose process and the Blandford--Znajek mechanism.

The angular velocity of the event horizon is defined by
\begin{equation}
\Omega_H
=
-\left.
\frac{g_{t\phi}}{g_{\phi\phi}}
\right|_{r=r_H}.
\label{eq:omega_H_definition}
\end{equation}

For the metric in Eq.~\eqref{eq:rotating_KR_metric}, this gives
\begin{equation}
\Omega_H^{\rm KR}
=
\frac{a}{r_H^2+a^2}.
\label{eq:omega_H_KR}
\end{equation}

Although this expression has the same formal structure as in the Kerr spacetime, the horizon radius \(r_H\) is different because it is determined by the modified horizon equation \eqref{eq:horizon_equation}. Therefore, the KR hair amplitude affects the horizon angular velocity indirectly through the modified horizon structure:
\begin{equation}
\Omega_H^{\rm KR}
=
\Omega_H^{\rm KR}(M,a,\Gamma,s).
\label{eq:omega_H_dependence}
\end{equation}

In the Kerr limit, \(\Gamma=0\), one obtains
\begin{equation}
\Omega_H^{\rm Kerr}
=
\frac{a}{\left(r_+^{\rm Kerr}\right)^2+a^2}.
\label{eq:omega_H_kerr}
\end{equation}

The difference between \(\Omega_H^{\rm KR}\) and \(\Omega_H^{\rm Kerr}\) controls the purely geometrical part of the leading fixed-flux BZ comparison. Whether this change survives in the physical jet power depends additionally on which magnetic quantity is held fixed and on the global magnetospheric solution. We therefore use \(\Omega_H^{\rm KR}\) below as one ingredient of the BZ model, rather than interpreting it alone as an observable efficiency change.

\section{Electromagnetic field and force-free magnetosphere}
\label{sec:force_free_magnetosphere}

We next summarize the electromagnetic setup. The black hole is assumed to be embedded in a stationary, axisymmetric, force-free magnetosphere, as appropriate for the leading BZ treatment. Plasma inertia is neglected relative to the electromagnetic energy density, leaving the field and the background geometry to determine the leading dynamics.

The Kalb--Ramond hair amplitude \(\Gamma\) enters the electromagnetic problem through the adopted background geometry, changing the horizon radius \(r_H\) and hence \(\Omega_H^{\rm KR}\). The actual BZ response, however, is jointly controlled by the horizon angular velocity, magnetic flux, and force-free field structure. Thus a geometrical change in \(\Omega_H\) need not translate into the same change in jet power under every magnetic prescription.

\subsection{Stationary axisymmetric electromagnetic field}
\label{subsec:stationary_axisymmetric_EM}

The electromagnetic field tensor is defined in terms of the vector potential \(A_\mu\) as
\begin{equation}
F_{\mu\nu}
=
\partial_\mu A_\nu
-
\partial_\nu A_\mu .
\label{eq:EM_tensor}
\end{equation}

In a stationary and axisymmetric spacetime, we assume that the electromagnetic field also respects the same symmetries. Therefore,
\begin{equation}
\partial_t A_\mu=0,
\qquad
\partial_\phi A_\mu=0 .
\label{eq:stationary_axisymmetric_condition}
\end{equation}

The relevant Maxwell equations in curved spacetime are
\begin{equation}
\nabla_\nu F^{\mu\nu}
=
J^\mu ,
\label{eq:maxwell_inhomogeneous}
\end{equation}
and
\begin{equation}
\nabla_{[\alpha}F_{\mu\nu]}=0 ,
\label{eq:maxwell_homogeneous}
\end{equation}
where \(J^\mu\) is the electromagnetic four-current. Equivalently, Eq.~\eqref{eq:maxwell_homogeneous} follows from the definition of \(F_{\mu\nu}\) in Eq.~\eqref{eq:EM_tensor}.

For a stationary and axisymmetric magnetosphere, the azimuthal component of the vector potential,
\begin{equation}
\Psi \equiv A_\phi ,
\label{eq:magnetic_flux_function}
\end{equation}
can be interpreted as a magnetic flux function. The surfaces of constant \(\Psi\) represent magnetic surfaces, and magnetic field lines lie on these surfaces.

The magnetic flux through a surface \(\mathcal{S}\) is given by
\begin{equation}
\Phi_B
=
\int_{\mathcal{S}} F_{\theta\phi}\,d\theta d\phi .
\label{eq:magnetic_flux_general}
\end{equation}

For an axisymmetric field configuration, this can be written as
\begin{equation}
\Phi_B
=
2\pi
\left[
A_\phi(\theta_2)-A_\phi(\theta_1)
\right].
\label{eq:magnetic_flux_Aphi}
\end{equation}

For an axisymmetric split-monopole topology, $A_\phi$ is proportional to $1-\cos\theta$, but its normalization should be tied to the magnetic flux rather than to a coordinate area. We therefore leave the normalization generic at this stage and define
\begin{equation}
A_\phi
=
\frac{\Phi_H}{2\pi}(1-\cos\theta),
\label{eq:split_monopole_flux}
\end{equation}
so that the flux through one hemisphere is $\Phi_H$. The relation between $\Phi_H$, a locally measured horizon field, and the reduced coordinate-radius proxy $2\pi B_0r_H^2$ is treated explicitly in Sec.~\ref{subsec:BZ_jet_power}. This distinction avoids identifying the phenomenological $r_H^2$ normalization with an invariant physical horizon flux.

\subsection{Force-free condition}
\label{subsec:force_free_condition}

The force-free approximation is the standard analytical limit for a magnetically dominated black-hole magnetosphere \cite{Blandford1977,Komissarov2001,McKinney2006}. Stationary and axisymmetric force-free solutions, including their field-line rotation and regularity properties, have been developed further in Refs.~\cite{McKinneyNarayan2007,Penna2015,PanYu2015}, while time-dependent numerical work tests the regime beyond idealized stationary configurations \cite{Parfrey2019}. The force-free condition is
\begin{equation}
F_{\mu\nu}J^\nu=0 .
\label{eq:force_free_condition}
\end{equation}

This condition means that the electromagnetic field dominates the plasma inertia and that the plasma adjusts itself so that the Lorentz force is zero. In addition, the force-free regime requires the electromagnetic field to be magnetically dominated,
\begin{equation}
B^2-E^2>0 ,
\label{eq:magnetically_dominated}
\end{equation}
and degenerate,
\begin{equation}
E\cdot B=0 .
\label{eq:degenerate_condition}
\end{equation}

In covariant form, the degeneracy condition can be expressed as
\begin{equation}
F_{\mu\nu}\,{}^\star F^{\mu\nu}=0 ,
\label{eq:covariant_degeneracy}
\end{equation}
where \({}^\star F^{\mu\nu}\) is the dual electromagnetic tensor. These conditions are standard assumptions in the analytical treatment of Blandford--Znajek magnetospheres.

\subsection{Magnetic field-line angular velocity}
\label{subsec:field_line_angular_velocity}

In a stationary and axisymmetric force-free magnetosphere, magnetic field lines rotate with an angular velocity \(\Omega_F\). This quantity is constant along each magnetic surface and can be written as a function of the flux function:
\begin{equation}
\Omega_F=\Omega_F(A_\phi).
\label{eq:omegaF_flux_function}
\end{equation}

The field-line angular velocity, a conserved quantity on a magnetic surface in stationary axisymmetric force-free electrodynamics \cite{Blandford1977,McKinneyNarayan2007,Penna2015}, is defined by
\begin{equation}
\Omega_F
=
-\frac{F_{tr}}{F_{\phi r}}
=
-\frac{F_{t\theta}}{F_{\phi\theta}} .
\label{eq:omegaF_definition}
\end{equation}

This expression follows from the degeneracy condition of the force-free electromagnetic field. Physically, \(\Omega_F\) measures how fast the magnetic field lines rotate with respect to an observer at infinity.

For outward electromagnetic extraction from a prograde rotating horizon, the standard BZ condition is that the field lines rotate in the same sense as, but more slowly than, the horizon \cite{Blandford1977,Penna2015,Noda2020}. Therefore,
\begin{equation}
0<\Omega_F<\Omega_H^{\rm KR}.
\label{eq:energy_extraction_condition}
\end{equation}

If this condition is satisfied, the rotating black hole can transfer part of its rotational energy to the electromagnetic field, producing an outgoing Poynting flux. In the present work, the horizon angular velocity is given by
\begin{equation}
\Omega_H^{\rm KR}
=
\frac{a}{r_H^2+a^2},
\label{eq:omegaH_KR_again}
\end{equation}
where \(r_H\) is determined by the modified horizon equation
\begin{equation}
r_H^2-2Mr_H+a^2+\Gamma r_H^n=0 .
\label{eq:horizon_gamma_again}
\end{equation}

Thus, the KR hair amplitude affects the energy extraction condition through the modified value of \(\Omega_H^{\rm KR}\).

\subsection{Electromagnetic energy and angular momentum fluxes}
\label{subsec:EM_fluxes}

The energy and angular momentum carried by the electromagnetic field are described by the electromagnetic stress-energy tensor,
\begin{equation}
T^{\mu\nu}_{\rm EM}
=
F^{\mu\alpha}F^\nu{}_{\alpha}
-
\frac{1}{4}g^{\mu\nu}F_{\alpha\beta}F^{\alpha\beta}.
\label{eq:EM_stress_energy}
\end{equation}

The radial flux of electromagnetic energy is
\begin{equation}
\mathcal{E}^r
=
-T^r{}_{t},
\label{eq:energy_flux_density}
\end{equation}
while the radial flux of angular momentum is
\begin{equation}
\mathcal{L}^r
=
T^r{}_{\phi}.
\label{eq:angular_momentum_flux_density}
\end{equation}

For stationary and axisymmetric force-free fields, the conserved energy and angular-momentum fluxes satisfy the standard relation \cite{Blandford1977,McKinneyNarayan2007}
\begin{equation}
\mathcal{E}^r
=
\Omega_F \mathcal{L}^r .
\label{eq:energy_angular_momentum_relation}
\end{equation}

The total electromagnetic power extracted from the black hole horizon is obtained by integrating the energy flux over the horizon surface:
\begin{equation}
P_{\rm EM}
=
-\int_H T^r{}_{t}\sqrt{-g}\,d\theta d\phi .
\label{eq:total_EM_power}
\end{equation}

The sign convention is chosen such that \(P_{\rm EM}>0\) corresponds to outward energy extraction from the black hole. This is the physical basis of the Blandford--Znajek process in the rotating Kalb--Ramond background.

\subsection{Energy extraction condition}
\label{subsec:energy_extraction_condition}

After imposing horizon regularity, the BZ energy flux contains the characteristic factor \(\Omega_F(\Omega_H-\Omega_F)\) \cite{Blandford1977,NarayanPennaSadowski2013,Penna2015}. In the KR background this gives
\begin{equation}
P_{\rm EM}
\propto
\Omega_F
\left(
\Omega_H^{\rm KR}
-
\Omega_F
\right)
\Phi_H^2 .
\label{eq:BZ_condition_form}
\end{equation}

Therefore, outward energy extraction requires
\begin{equation}
\Omega_F
\left(
\Omega_H^{\rm KR}
-
\Omega_F
\right)>0 .
\label{eq:positive_power_condition}
\end{equation}

For a rotating black hole with \(\Omega_H^{\rm KR}>0\), this gives
\begin{equation}
0<\Omega_F<\Omega_H^{\rm KR}.
\label{eq:final_energy_condition}
\end{equation}

For the usual impedance-matched/split-monopole benchmark, the power is maximized near \(\Omega_F/\Omega_H\simeq 1/2\) \cite{NarayanPennaSadowski2013,Penna2015},
\begin{equation}
\Omega_F
\simeq
\frac{1}{2}\Omega_H^{\rm KR}.
\label{eq:optimal_omegaF}
\end{equation}

This benchmark condition will be used in the next section to construct the leading BZ power. The KR hair amplitude changes \(\Omega_H^{\rm KR}\) through the horizon geometry, but the resulting change in total power remains conditional on the adopted magnetic-flux prescription.

\section{Blandford--Znajek power in the Kalb--Ramond background}
\label{sec:BZ_power_KR}

We now apply the leading Blandford--Znajek jet-power scaling to the rotating Kalb--Ramond background. A rotating black hole immersed in a large-scale magnetic field can transfer rotational energy to an outgoing electromagnetic Poynting flux. Our treatment is intentionally factorized: we use the standard stationary, axisymmetric force-free BZ dependence on horizon flux and field-line rotation, but we do not solve a KR-specific Grad--Shafranov problem or determine a new global force-free magnetosphere from first principles. Accordingly, \(\kappa_{\rm BZ}\) and the benchmark ratio \(\Omega_F/\Omega_H\) are treated as model inputs, and the results below should be read as leading-order geometric sensitivity tests \cite{Armas2020,Komissarov2009,Camilloni2024}.

\subsection{General expression for electromagnetic energy flux}
\label{subsec:general_EM_flux}

The electromagnetic energy flux through the event horizon is defined by
\begin{equation}
P_{\rm EM}
=
-\int_H T^r{}_{t}\sqrt{-g}\,d\theta d\phi ,
\label{eq:EM_power_horizon}
\end{equation}
where \(T^\mu{}_\nu\) is the electromagnetic stress-energy tensor. The minus sign is chosen so that positive \(P_{\rm EM}\) corresponds to outward energy extraction from the black hole.

For a stationary and axisymmetric force-free magnetosphere, the leading BZ scaling can be written in terms of the field-line angular velocity \(\Omega_F\), the horizon angular velocity \(\Omega_H^{\rm KR}\), and the magnetic flux threading the horizon \cite{Blandford1977,Tchekhovskoy2010,NarayanPennaSadowski2013,Camilloni2024}. In a compact convention in which numerical and geometric factors are absorbed into \(\kappa_{\rm BZ}\), we write
\begin{equation}
P_{\rm BZ}^{\rm KR}
=
\kappa_{\rm BZ}
\Phi_H^2
\Omega_F
\left(
\Omega_H^{\rm KR}
-
\Omega_F
\right),
\label{eq:BZ_power_general_KR}
\end{equation}
where \(\kappa_{\rm BZ}\) is a dimensionless coefficient depending on the magnetic-field geometry. The quantity \(\Phi_H\) denotes the magnetic flux through one horizon hemisphere, with its invariant definition specified below.

It is useful to introduce the dimensionless ratio
\begin{equation}
x
=
\frac{\Omega_F}{\Omega_H^{\rm KR}} .
\label{eq:x_definition}
\end{equation}
Then Eq.~\eqref{eq:BZ_power_general_KR} becomes
\begin{equation}
P_{\rm BZ}^{\rm KR}
=
\kappa_{\rm BZ}
\Phi_H^2
\left(
\Omega_H^{\rm KR}
\right)^2
x(1-x).
\label{eq:BZ_power_x}
\end{equation}

This expression clearly shows that the energy extraction is possible only when
\begin{equation}
0<x<1,
\label{eq:x_condition}
\end{equation}
or equivalently
\begin{equation}
0<\Omega_F<\Omega_H^{\rm KR}.
\label{eq:BZ_energy_condition}
\end{equation}

Therefore, the magnetic field lines must rotate more slowly than the black hole horizon. If \(\Omega_F=0\), the field lines do not rotate and no electromagnetic power is extracted. If \(\Omega_F=\Omega_H^{\rm KR}\), the field lines corotate with the horizon and again the net extracted power vanishes. The energy extraction is efficient only between these two limiting cases.

\subsection{Blandford--Znajek jet power}
\label{subsec:BZ_jet_power}

The horizon angular velocity of the rotating Kalb--Ramond black hole is
\begin{equation}
\Omega_H^{\rm KR}
=
\frac{a}{r_H^2+a^2},
\label{eq:omegaH_KR_BZ}
\end{equation}
where the horizon radius \(r_H\) is determined by
\begin{equation}
r_H^2-2Mr_H+a^2+\Gamma r_H^n=0 .
\label{eq:horizon_KR_BZ}
\end{equation}

Thus, although Eq.~\eqref{eq:omegaH_KR_BZ} has the same formal structure as the Kerr expression, the value of \(r_H\) is modified by the Kalb--Ramond hair amplitude \(\Gamma\). As a result, the jet power becomes
\begin{equation}
P_{\rm BZ}^{\rm KR}
=
\kappa_{\rm BZ}
\Phi_H^2
\left[
\frac{a}{r_H^2+a^2}
\right]^2
x(1-x).
\label{eq:BZ_power_KR_explicit}
\end{equation}

The magnetic flux entering the BZ formula must be defined invariantly. On a spatial section of the KR horizon, the induced two-metric gives
\begin{equation}
\sqrt{g_{\theta\theta}g_{\phi\phi}}\bigg|_{r_H}
=
\left(r_H^2+a^2\right)\sin\theta,
\label{eq:horizon_area_element}
\end{equation}
where the horizon identity $2Mr_H-\Gamma r_H^n=r_H^2+a^2$ has been used. Therefore the proper magnetic flux through one hemisphere is
\begin{equation}
\Phi_H
=
\int_0^{2\pi}\!d\phi\int_0^{\pi/2}\!d\theta\,
B_H^{\hat r}\left(r_H^2+a^2\right)\sin\theta,
\label{eq:proper_horizon_flux_general}
\end{equation}
with $B_H^{\hat r}$ the locally measured normal magnetic field. This is the geometrical quantity that appears in the standard BZ flux definition \cite{Tchekhovskoy2011,NarayanPennaSadowski2013}. For an approximately uniform local normal field $B_H$ on the horizon cross-section,
\begin{equation}
\Phi_H
\simeq
2\pi B_H\left(r_H^2+a^2\right).
\label{eq:proper_horizon_flux_uniform}
\end{equation}

Three magnetic prescriptions will be compared. First, one may compare spacetimes at fixed total flux $\Phi_H$, in which case all leading geometrical dependence enters through $\Omega_H$. Second, one may hold the local normal field $B_H$ fixed and use the proper flux in Eq.~\eqref{eq:proper_horizon_flux_uniform}. Third, for comparison with the earlier phenomenological parameterization, we define the reduced proxy
\begin{equation}
\Phi_{\rm proxy}
\equiv
2\pi B_0 r_H^2,
\label{eq:flux_BZ_KR}
\end{equation}
which should not be identified with the invariant flux of a uniform physical horizon field. Because the magnetic-flux prescription is astrophysical rather than purely geometrical, the numerical analysis below presents the fixed-$\Phi_H$ case as the baseline and uses Eq.~\eqref{eq:flux_BZ_KR} only for a sensitivity comparison.

\subsection{Optimal energy extraction condition}
\label{subsec:optimal_BZ_condition}

The factor
\begin{equation}
x(1-x)
\label{eq:x_factor}
\end{equation}
controls the efficiency of electromagnetic energy extraction. Its maximum is obtained from
\begin{equation}
\frac{d}{dx}
\left[
x(1-x)
\right]
=
1-2x=0 .
\label{eq:derivative_x}
\end{equation}

Thus, the optimal value is
\begin{equation}
x=\frac{1}{2},
\label{eq:x_optimal}
\end{equation}
or
\begin{equation}
\Omega_F
=
\frac{1}{2}\Omega_H^{\rm KR}.
\label{eq:omegaF_optimal_KR}
\end{equation}

Under this condition, the Blandford--Znajek power becomes
\begin{equation}
P_{\rm BZ,opt}^{\rm KR}
=
\frac{1}{4}
\kappa_{\rm BZ}
\Phi_H^2
\left(
\Omega_H^{\rm KR}
\right)^2 .
\label{eq:BZ_power_optimal_KR}
\end{equation}

For a fixed local normal field $B_H$, Eqs.~\eqref{eq:BZ_power_optimal_KR} and \eqref{eq:proper_horizon_flux_uniform} give
\begin{equation}
P_{\rm BZ,opt}^{\rm KR}
\simeq
\pi^2\kappa_{\rm BZ}B_H^2
\left(r_H^2+a^2\right)^2
\left[\frac{a}{r_H^2+a^2}\right]^2
=
\pi^2\kappa_{\rm BZ}B_H^2a^2.
\label{eq:BZ_power_fixed_local_B}
\end{equation}
Thus, within the leading-order factorized BZ model and for the same local horizon field, the explicit KR dependence cancels at fixed $a$. Residual dependence can still arise in a full force-free solution through changes in field geometry, the coefficient $\kappa_{\rm BZ}$, plasma loading, or the global magnetosphere \cite{Camilloni2024,NarayanPennaSadowski2013}.

For later numerical analysis, we introduce
\begin{equation}
a_*=\frac{a}{M},
\qquad
R_H=\frac{r_H}{M},
\label{eq:dimensionless_spin_horizon}
\end{equation}
so that
\begin{equation}
M\Omega_H^{\rm KR}
=
\frac{a_*}{R_H^2+a_*^2}.
\label{eq:dimensionless_omegaH_KR}
\end{equation}
For the fixed-flux baseline, a convenient reduced power is
\begin{equation}
\widetilde{\mathcal P}_{\Phi}^{\rm KR}
\equiv
\frac{4M^2P_{\rm BZ,opt}^{\rm KR}}{\kappa_{\rm BZ}\Phi_H^2}
=
\left(M\Omega_H^{\rm KR}\right)^2
=
\left[\frac{a_*}{R_H^2+a_*^2}\right]^2.
\label{eq:normalized_BZ_power_final}
\end{equation}
This normalization isolates the geometrical dependence without assigning a local magnetic-field strength.

\subsection{Kerr limit}
\label{subsec:Kerr_limit_BZ}

A physically consistent result must reduce to the standard Kerr case when the Kalb--Ramond correction vanishes. In the limit
\begin{equation}
\Gamma\rightarrow 0,
\label{eq:gamma_zero}
\end{equation}
the horizon equation becomes
\begin{equation}
r_H^2-2Mr_H+a^2=0,
\label{eq:horizon_kerr_limit}
\end{equation}
and the outer horizon reduces to
\begin{equation}
r_H\rightarrow r_+^{\rm Kerr}
=
M+\sqrt{M^2-a^2}.
\label{eq:rplus_kerr_limit}
\end{equation}

Consequently, the horizon angular velocity becomes
\begin{equation}
\Omega_H^{\rm KR}
\rightarrow
\Omega_H^{\rm Kerr}
=
\frac{a}{\left(r_+^{\rm Kerr}\right)^2+a^2}.
\label{eq:omegaH_kerr_limit}
\end{equation}

Therefore,
\begin{equation}
P_{\rm BZ}^{\rm KR}
\rightarrow
P_{\rm BZ}^{\rm Kerr}.
\label{eq:BZ_kerr_limit}
\end{equation}

This confirms that the rotating Kalb--Ramond result continuously recovers the standard Kerr Blandford--Znajek power in the appropriate limit.

\subsection{Deviation from the Kerr jet power}
\label{subsec:deviation_Kerr}

To quantify the effect of the KR hair amplitude, we define the relative deviation of the jet power from the Kerr value:
\begin{equation}
\delta_{\rm BZ}
=
\frac{
P_{\rm BZ}^{\rm KR}
-
P_{\rm BZ}^{\rm Kerr}
}{
P_{\rm BZ}^{\rm Kerr}
}.
\label{eq:delta_BZ_power}
\end{equation}

If the same magnetic flux and the same value of \(x=\Omega_F/\Omega_H\) are assumed in both spacetimes, the ratio of the jet powers becomes
\begin{equation}
\frac{
P_{\rm BZ}^{\rm KR}
}{
P_{\rm BZ}^{\rm Kerr}
}
=
\left(
\frac{
\Omega_H^{\rm KR}
}{
\Omega_H^{\rm Kerr}
}
\right)^2 .
\label{eq:BZ_ratio_same_flux}
\end{equation}

For a fixed locally measured normal field $B_H$, the proper-flux expression in Eq.~\eqref{eq:BZ_power_fixed_local_B} instead gives
\begin{equation}
\left.\frac{P_{\rm BZ}^{\rm KR}}{P_{\rm BZ}^{\rm Kerr}}\right|_{B_H}
=1
\qquad
\text{(same $a$, $B_H$, $x$, and $\kappa_{\rm BZ}$)},
\label{eq:BZ_ratio_fixed_local_B}
\end{equation}
at this leading order. By contrast, the reduced proxy $\Phi_{\rm proxy}\propto r_H^2$ gives
\begin{equation}
\left.\frac{P_{\rm BZ}^{\rm KR}}{P_{\rm BZ}^{\rm Kerr}}\right|_{\rm proxy}
=
\left(\frac{r_H^{\rm KR}}{r_+^{\rm Kerr}}\right)^4
\left(\frac{\Omega_H^{\rm KR}}{\Omega_H^{\rm Kerr}}\right)^2.
\label{eq:BZ_ratio_fixed_B0}
\end{equation}
The three prescriptions therefore need not give the same qualitative dependence on $\bar{\Gamma}$. This is not a contradiction: $\Phi_H$, a local normal field, and a coordinate-radius proxy correspond to different astrophysical assumptions. The fixed-$\Phi_H$ case is used below as the baseline because it follows directly from the invariant BZ scaling, while the other two cases are used to expose model dependence.

In the following section, we numerically analyze \(r_H\), \(\Omega_H^{\rm KR}\), and \(P_{\rm BZ}^{\rm KR}\) as functions of the spin parameter \(a_*\) and the Kalb--Ramond hair amplitude \(\Gamma\).

\section{Numerical analysis}
\label{sec:numerical_analysis}

We first examine numerically how the KR hair changes the horizon radius, the horizon angular velocity, and the leading BZ power. These calculations are not tied to a particular source; their purpose is to isolate the geometrical response before introducing the observational inputs used in the Bayesian analysis.

For numerical convenience, we introduce
\begin{equation}
a_* = \frac{a}{M},
\qquad
R_H = \frac{r_H}{M},
\label{eq:dimensionless_parameters_numerical}
\end{equation}
and use the primary $s=3/2$ dimensionless hair amplitude
\begin{equation}
\bar{\Gamma}=\frac{\Gamma}{M^{4/3}}.
\label{eq:dimensionless_gamma}
\end{equation}
The horizon equation becomes
\begin{equation}
R_H^2-2R_H+a_*^2+\bar{\Gamma}R_H^{2/3}=0,
\label{eq:dimensionless_horizon_equation}
\end{equation}
and the outer horizon is the largest positive real root. Only parameter values with a regular positive outer horizon are retained.

For the representative curves we use
\begin{equation}
\bar{\Gamma}=\{-0.10,-0.05,0,0.02,0.04\},
\label{eq:gamma_values_numerical}
\end{equation}
with spin values restricted to the horizon-admitting region. The Kerr case is $\bar{\Gamma}=0$.

\begin{figure}[!htbp]
    \centering
    \includegraphics[width=0.72\linewidth]{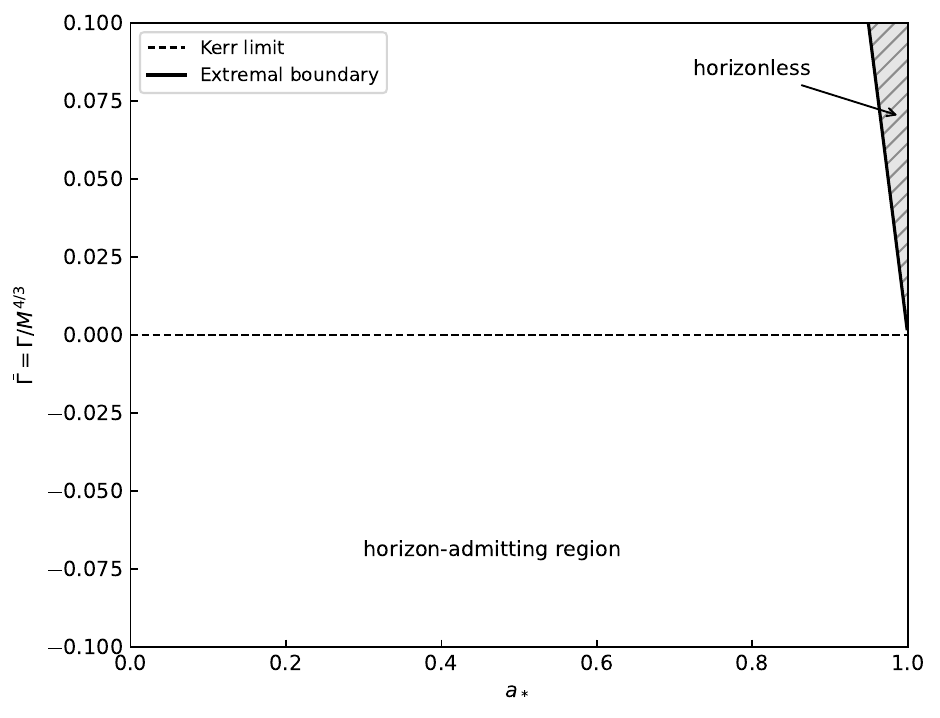}
    \caption{Horizon-admitting parameter space in the $(a_*,\bar{\Gamma})$ plane for the primary $s=3/2$ ($n=2/3$) benchmark, shown over the numerical range $-0.10\leq\bar{\Gamma}\leq0.10$. The solid curve is the extremal boundary from Eq.~\eqref{eq:s15_extremal_boundary}. The hatched wedge above and to the right of this curve is horizonless and is excluded, whereas points below the boundary admit a positive outer horizon. The dashed horizontal line denotes the Kerr limit $\bar{\Gamma}=0$.}
    \label{fig:parameter_space}
\end{figure}

The extremal boundary for the primary $s=3/2$ benchmark can be written parametrically. If $R_e$ is the degenerate horizon radius, simultaneous solution of $f(R_e)=f'(R_e)=0$ for
$f(R)=R^2-2R+a_*^2+\bar{\Gamma}R^{2/3}$ gives
\begin{equation}
a_*^2=R_e(2R_e-1),
\qquad
\bar{\Gamma}_{\rm ext}=3(1-R_e)R_e^{1/3}.
\label{eq:s15_extremal_boundary}
\end{equation}
Figure~\ref{fig:parameter_space} displays the corresponding allowed domain. The shaded high-spin, positive-$\bar{\Gamma}$ wedge is horizonless and is excluded in all subsequent calculations.

\subsection{Horizon radius and angular velocity}
\label{subsec:horizon_radius_angular_velocity}

\begin{figure}[!htbp]
    \centering
    \includegraphics[width=0.72\linewidth]{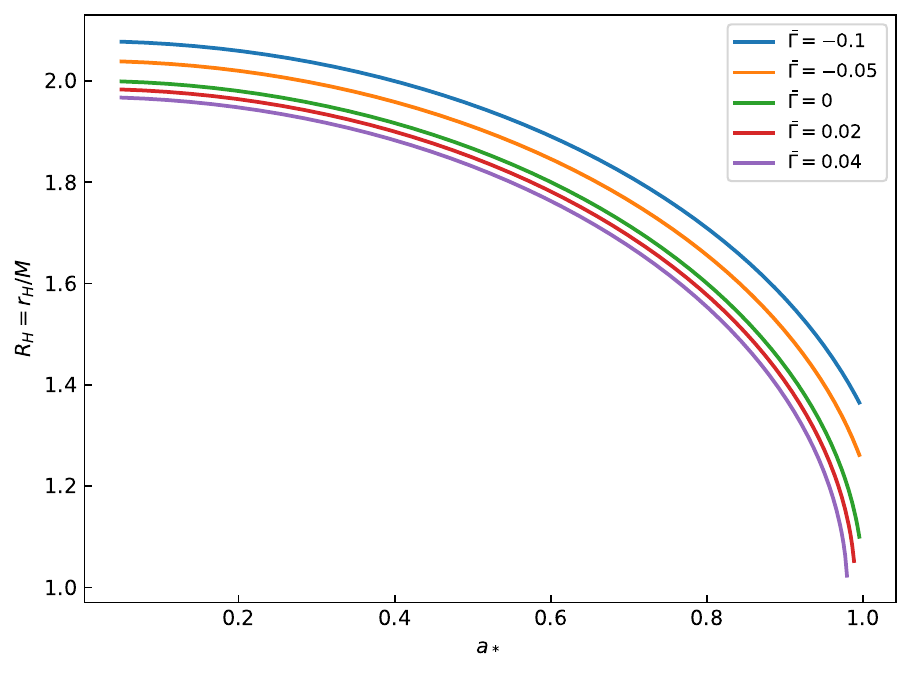}
    \caption{Dimensionless outer horizon radius \(R_H=r_H/M\) as a function of the spin parameter \(a_*\) for different values of the dimensionless hair amplitude \(\bar{\Gamma}\). The curve with \(\bar{\Gamma}=0\) corresponds to the Kerr limit.}
    \label{fig:horizon_radius}
\end{figure}

The dependence of the outer horizon radius on the spin parameter is shown in Fig.~\ref{fig:horizon_radius}. For all considered values of \(\bar{\Gamma}\), the horizon radius decreases as the spin increases. This behavior is similar to the Kerr case. However, the KR hair amplitude shifts the horizon location. Negative values of \(\bar{\Gamma}\) increase the horizon radius, whereas positive values reduce it relative to the Kerr black hole.

The dimensionless horizon angular velocity is given by
\begin{equation}
M\Omega_H^{\rm KR}
=
\frac{a_*}{R_H^2+a_*^2} .
\label{eq:dimensionless_omegaH_numerical}
\end{equation}
Although this expression has the same formal structure as in the Kerr spacetime, the horizon radius \(R_H\) is modified by \(\bar{\Gamma}\). Therefore, the KR hair amplitude affects \(M\Omega_H^{\rm KR}\) indirectly through the modified horizon structure.

\begin{figure}[!htbp]
    \centering
    \includegraphics[width=0.98\linewidth]{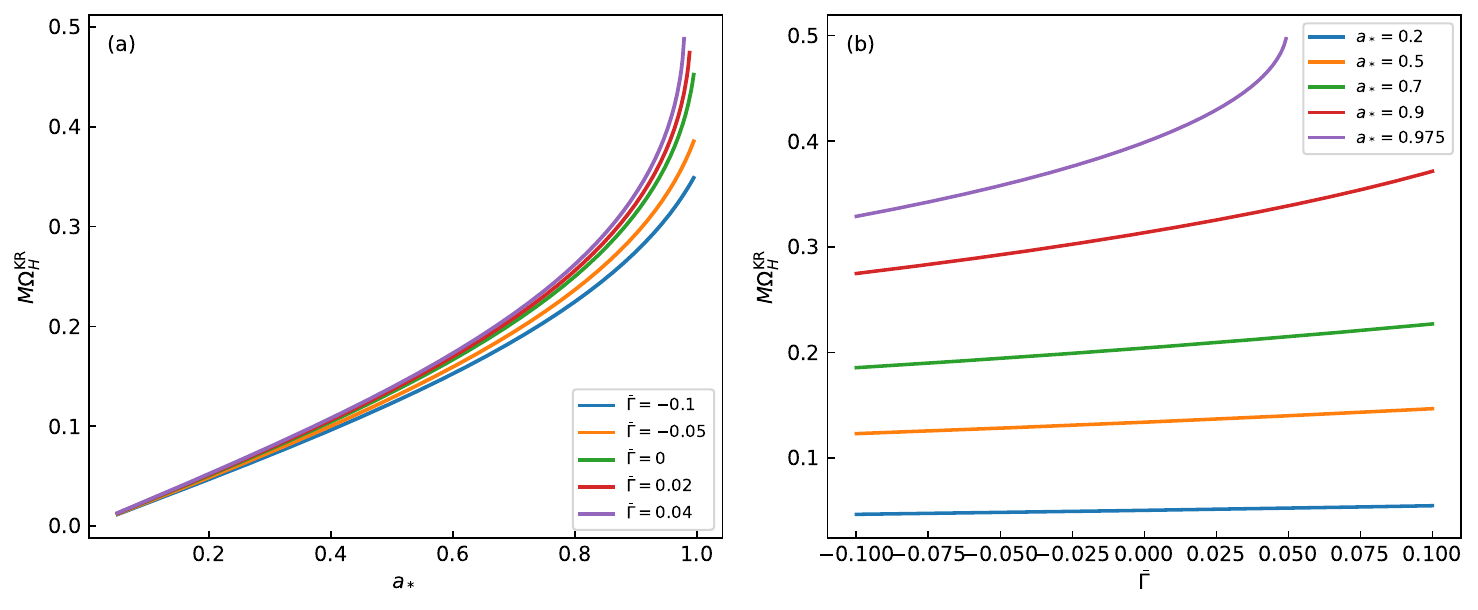}
    \caption{Dimensionless horizon angular velocity $M\Omega_H^{\rm KR}$. Panel~(a): dependence on the spin parameter $a_*$ for representative values of $\bar{\Gamma}$. Panel~(b): dependence on $\bar{\Gamma}$ for selected fixed spins. Positive values of $\bar{\Gamma}$ enhance the horizon angular velocity, while negative values suppress it.}
    \label{fig:horizon_angular_velocity}
\end{figure}

Figure~\ref{fig:horizon_angular_velocity} shows the behavior of the horizon angular velocity in two complementary ways. Panel~(a) presents $M\Omega_H^{\rm KR}$ as a function of the spin parameter for representative values of $\bar{\Gamma}$, while panel~(b) shows its dependence on $\bar{\Gamma}$ for several fixed spins. In both panels the same conclusion emerges: positive $\bar{\Gamma}$ gives a smaller horizon radius and therefore a larger horizon angular velocity, whereas negative $\bar{\Gamma}$ enlarges the horizon and suppresses $M\Omega_H^{\rm KR}$.

\subsection{Jet power as a function of spin}
\label{subsec:jet_power_spin}

Under the optimal Blandford--Znajek condition,
\begin{equation}
\Omega_F=\frac{1}{2}\Omega_H^{\rm KR},
\label{eq:optimal_condition_numerical}
\end{equation}
we adopt the invariant fixed-flux baseline introduced in Eq.~\eqref{eq:normalized_BZ_power_final}. The reduced dimensionless power is therefore
\begin{equation}
\widetilde{\mathcal{P}}_{\Phi}^{\rm KR}
=
\left(M\Omega_H^{\rm KR}\right)^2
=
\left[
\frac{a_*}{R_H^2+a_*^2}
\right]^2 .
\label{eq:reduced_power_numerical}
\end{equation}
This quantity isolates the spacetime dependence at fixed total horizon flux and does not require a coordinate-based magnetic-field normalization.

\begin{figure}[!htbp]
    \centering
    \includegraphics[width=0.72\linewidth]{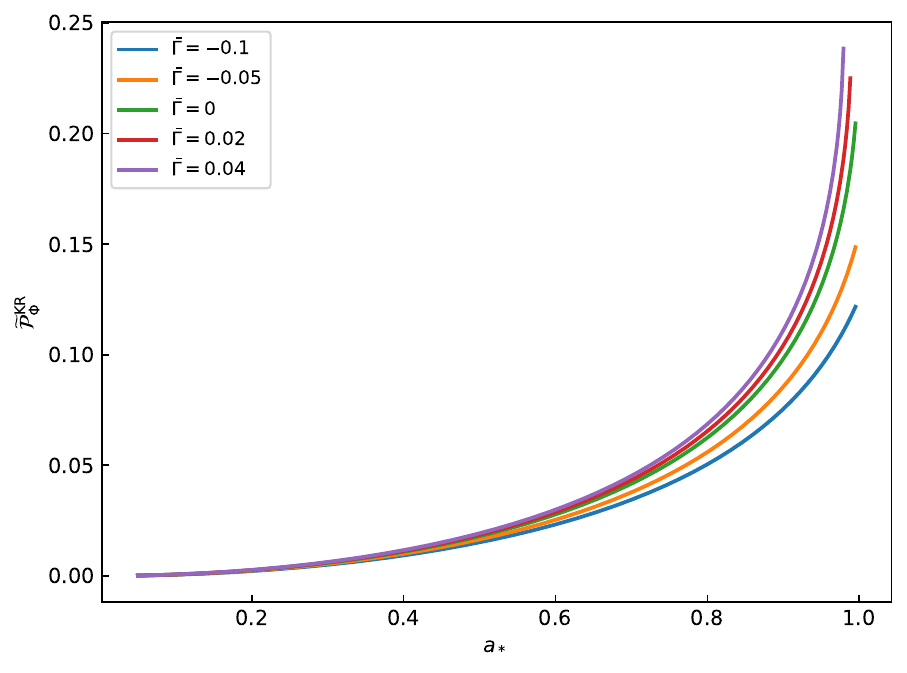}
    \caption{Fixed-flux reduced Blandford--Znajek power $\widetilde{\mathcal{P}}_{\Phi}^{\rm KR}=(M\Omega_H^{\rm KR})^2$ as a function of $a_*$. Positive $\bar{\Gamma}$ enhances the power and negative $\bar{\Gamma}$ suppresses it.}
    \label{fig:reduced_power}
\end{figure}

The behavior of $\widetilde{\mathcal{P}}_{\Phi}^{\rm KR}$ is shown in Fig.~\ref{fig:reduced_power}. The power increases strongly with spin. At fixed $a_*$, positive $\bar{\Gamma}$ decreases the horizon radius, increases $\Omega_H$, and therefore enhances the fixed-flux BZ power; negative $\bar{\Gamma}$ has the opposite effect.

\subsection{Comparison with the Kerr case}
\label{subsec:comparison_kerr}

For the fixed-flux baseline, the relative deviation is
\begin{equation}
\frac{P_{\rm BZ}^{\rm KR}}{P_{\rm BZ}^{\rm Kerr}}
=
\left(\frac{\Omega_H^{\rm KR}}{\Omega_H^{\rm Kerr}}\right)^2,
\label{eq:power_ratio_fixed_flux_numerical}
\end{equation}
where
\begin{equation}
R_+^{\rm Kerr}=1+\sqrt{1-a_*^2},
\qquad
M\Omega_H^{\rm Kerr}=\frac{a_*}{(R_+^{\rm Kerr})^2+a_*^2} .
\label{eq:kerr_quantities_numerical}
\end{equation}
We also define the percentage deviation
\begin{equation}
\Delta P_{\rm BZ}^{(\%)}=100\left(\frac{P_{\rm BZ}^{\rm KR}}{P_{\rm BZ}^{\rm Kerr}}-1\right)\%.
\label{eq:relative_power_difference_numerical}
\end{equation}

\begin{figure}[!htbp]
    \centering
    \includegraphics[width=0.98\linewidth]{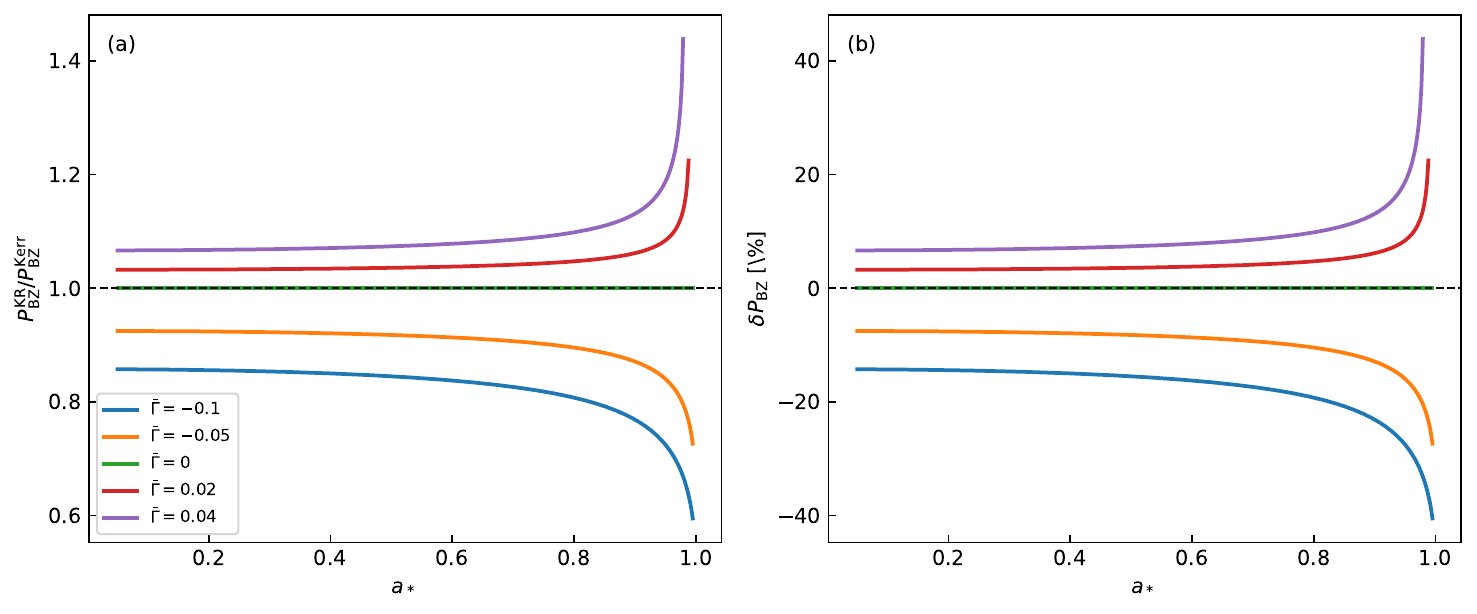}
    \caption{Fixed-flux deviation of the rotating Kalb--Ramond BZ power from Kerr. Panel~(a): $P_{\rm BZ}^{\rm KR}/P_{\rm BZ}^{\rm Kerr}$. Panel~(b): percentage deviation $\Delta P_{\rm BZ}^{(\%)}$. The horizontal reference lines denote the Kerr values.}
    \label{fig:power_ratio}
\end{figure}

Figure~\ref{fig:power_ratio} displays both forms. Negative $\bar{\Gamma}$ suppresses the fixed-flux power, while positive $\bar{\Gamma}$ enhances it. The effect grows toward high spin because the horizon angular velocity becomes increasingly sensitive to the horizon location.

\begin{figure}[!htbp]
    \centering
    \includegraphics[width=0.75\linewidth]{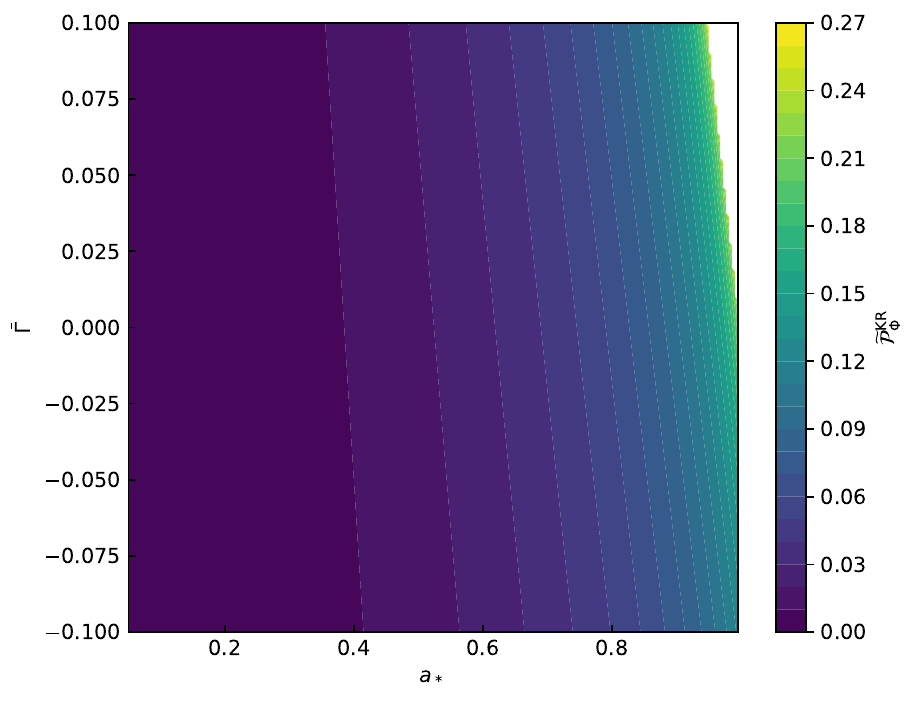}
    \caption{Contour plot of the fixed-flux reduced BZ power $\widetilde{\mathcal{P}}_{\Phi}^{\rm KR}$ in the $(a_*,\bar{\Gamma})$ plane for the primary $s=3/2$ ($n=2/3$) benchmark.}
    \label{fig:power_contour}
\end{figure}

Figure~\ref{fig:power_contour} shows the fixed-flux reduced power in the $(a_*,\bar{\Gamma})$ plane. Spin remains the dominant driver, but the KR deformation shifts the horizon angular velocity at fixed spin.

\subsection{Sensitivity to the magnetic-flux prescription}
\label{subsec:flux_prescription_audit}

\begin{figure}[!htbp]
    \centering
    \includegraphics[width=0.98\linewidth]{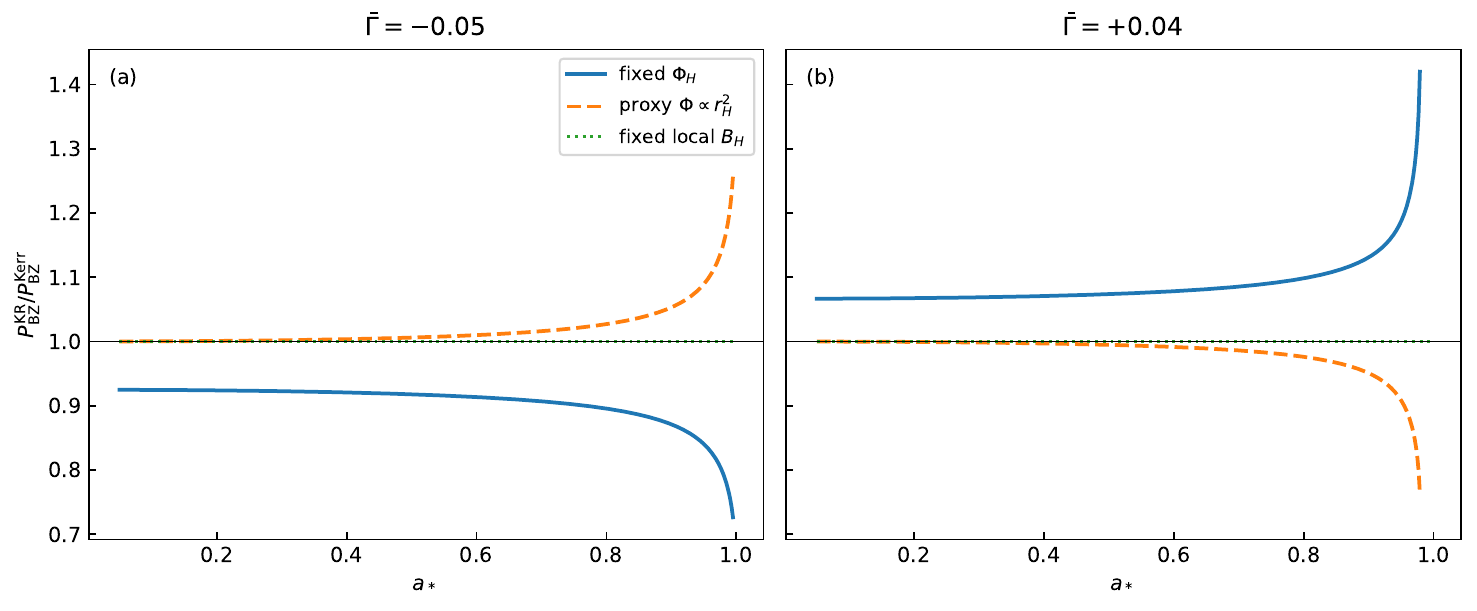}
    \caption{Sensitivity of the leading BZ power ratio to the magnetic-flux prescription for $\bar{\Gamma}=-0.05$ and $+0.04$. The fixed-$\Phi_H$ baseline follows the invariant BZ scaling, the dashed curve uses the reduced proxy $\Phi_{\rm proxy}\propto r_H^2$, and the dotted curve holds the local normal field $B_H$ fixed using the proper horizon area.}
    \label{fig:flux_prescription_audit}
\end{figure}

The leading BZ prediction is not determined by the spacetime geometry alone because one must also state which magnetic quantity is held fixed when two geometries are compared. Figure~\ref{fig:flux_prescription_audit} makes this dependence explicit. For fixed total horizon flux, positive $\bar{\Gamma}$ enhances the power and negative values suppress it. The phenomenological proxy $\Phi_{\rm proxy}\propto r_H^2$ produces the opposite trend because the radius-dependent proxy flux can dominate over the change in $\Omega_H$. Finally, if the locally measured normal field $B_H$ is held fixed and the proper horizon area is used, the leading-order ratio is unity, as predicted by Eq.~\eqref{eq:BZ_ratio_fixed_local_B}.

The differences between these prescriptions are not a small normalization effect. Any inference on $\bar{\Gamma}$ from jet power must specify what is being held fixed magnetically. In the Bayesian analysis we use fixed $\Phi_H$ as the baseline and reserve the other prescriptions for robustness checks.

\FloatBarrier
\section{Bayesian consistency test with jet-power proxies}
\label{sec:MCMC_constraints}

The numerical analysis shows that the predicted BZ response depends on the magnetic-flux prescription. We therefore use the invariant fixed-$\Phi_H$ case as the baseline for a Bayesian \emph{consistency test}, rather than interpreting the result immediately as a fundamental KR constraint. The jet-power proxy has substantial observational and modeling uncertainty, while the absolute BZ normalization depends on magnetic flux, field geometry, radiative conversion, and the empirical radio calibration. These quantities are absorbed into nuisance parameters in a generative likelihood \cite{Hogg2010,Trotta2017}.

We use GRO~J1655--40 and GRS~1915+105, two microquasars with well-studied X-ray and radio properties \cite{Zhang1997GRO,Remillard1999GRO,Gierlinski2001,Tomsick1999,Greiner2001,ZhangCuiChen2000,HjellmingRupen1995}. Jet--spin comparisons based on continuum-fitting spin measurements and radio proxies have been discussed extensively \cite{NarayanMcClintock2012,Steiner2012,McClintock2013,ZhangYu2015,Feng2017,vanVelzenFalcke2013,Daly2016,McKinneyORiordan2016}. The numerical proxy values adopted here follow the black-hole-binary sample used in non-Kerr BZ analyses \cite{Pei2016}. Because that proxy is already mass-normalized, black-hole masses do not enter the likelihood as independent parameters.

\begin{table}[!htbp]
\centering
\caption{Two-source data set used in the Bayesian consistency test. The spin measurements are included as Gaussian observational priors and the jet powers are treated as logarithmic proxies.}
\label{tab:two_source_data_revised}
\begin{tabular}{lccc}
\hline
Source & Spin prior & $P_{\rm jet}^{\rm proxy}$ & $\sigma_{\log P}$ \\
\hline
GRO~J1655--40 & $a_*=0.70\pm0.10$ & 70 & 0.35 \\
GRS~1915+105 & $a_*=0.975\pm0.015$ & 42 & 0.35 \\
\hline
\end{tabular}
\end{table}

\subsection{Generative model and priors}
\label{subsec:revised_generative_model}

For each source we write
\begin{equation}
P_{{\rm th},j}
=
K\,\widetilde{\mathcal P}_{\Phi}^{\rm KR}
\left(a_{*,j},\bar{\Gamma}\right),
\label{eq:revised_power_model}
\end{equation}
where $A\equiv\log_{10}K$ is a nuisance normalization. Under the fixed-flux baseline,
\begin{equation}
\widetilde{\mathcal P}_{\Phi}^{\rm KR}
=
\left[
\frac{a_*}{R_H^2+a_*^2}
\right]^2,
\label{eq:revised_reduced_power}
\end{equation}
with $R_H$ obtained numerically from the primary $s=3/2$ horizon equation
\begin{equation}
R_H^2-2R_H+a_*^2+\bar{\Gamma}R_H^{2/3}=0.
\label{eq:revised_horizon_equation}
\end{equation}
The logarithmic data model is
\begin{equation}
\log_{10}P_{{\rm jet},j}^{\rm obs}
\sim
\mathcal N\left[
A+\log_{10}\widetilde{\mathcal P}_{\Phi,j}^{\rm KR},
\sigma_{\log P,j}^2+\sigma_{\rm int}^2
\right],
\label{eq:revised_generative_distribution}
\end{equation}
with $\sigma_{\rm int}$ describing unmodeled source-to-source scatter. The full Gaussian normalization is retained because $\sigma_{\rm int}$ is sampled.

The sampled parameter vector is
\begin{equation}
\Theta=
\left(\bar{\Gamma},A,\sigma_{\rm int},a_*^{\rm GRO},a_*^{\rm GRS}\right).
\label{eq:revised_parameter_vector}
\end{equation}
The primary analysis assigns the physically transparent top-hat prior
\begin{equation}
\Pi_{\Gamma}^{\rm U}(\bar{\Gamma})=\mathcal U(-0.10,0.10),
\label{eq:gamma_uniform_prior}
\end{equation}
together with
\begin{equation}
0<A<4,
\qquad
0.02<\sigma_{\rm int}<0.60,
\label{eq:revised_uniform_priors}
\end{equation}
and the Gaussian spin-localization factors in Table~\ref{tab:two_source_data_revised}. Every sample must also possess a positive real outer horizon according to Eq.~\eqref{eq:revised_horizon_equation}. For positive $\bar{\Gamma}$, the boundary is equivalently described by Eq.~\eqref{eq:s15_extremal_boundary}; it is particularly restrictive for the near-extremal GRS~1915+105 spin prior.

To test whether the inferred deformation is controlled by the chosen prior, we repeat the complete calculation with a truncated-Gaussian sensitivity prior,
\begin{equation}
\Pi_{\Gamma}^{\rm G}(\bar{\Gamma})\propto
\exp\!\left[-\frac{(\bar{\Gamma}-\mu_{\Gamma})^2}{2\sigma_{\Gamma}^2}\right],
\qquad -0.10<\bar{\Gamma}<0.10,
\label{eq:gamma_gaussian_prior}
\end{equation}
with $\mu_{\Gamma}=-0.0144$ and $\sigma_{\Gamma}=0.0543$. These localization values are the mean and standard deviation of viable points from a preliminary broad scan over the same top-hat spin supports after requiring a horizon and $\chi^2_{\rm jet}<1$. They are used only as a controlled prior-sensitivity diagnostic and are not independent observational information. The jet likelihood, spin-localization factors, nuisance-parameter supports, horizon condition, and sampler settings are otherwise identical in the two runs.

The common $\bar{\Gamma}$ is only an effective phenomenological deformation shared by the two sources for this diagnostic exercise. Since $\Gamma$ is dimensional and a source-to-source mapping to a universal underlying KR coupling has not been established here, this parameter must not be interpreted as a direct universal bound on $\xi_2$, the KR vacuum expectation value, or the full theory. Moreover, the adopted spin estimates were obtained within Kerr-based spectral analyses. We use them as observational proxy priors to test information flow through the model, not as a self-consistent non-Kerr reanalysis of the X-ray spectra.
With only two jet-power data points, this five-parameter model is not expected to be likelihood-identifiable. In particular, the common normalization $K$, intrinsic scatter, two source spins, and the deformation amplitude can compensate one another. The common $K$ is itself a simplifying assumption because real systems need not carry the same horizon flux or radio-to-jet conversion. This underdetermination is deliberate: the purpose of the calculation is to measure whether the jet likelihood adds information after the priors are imposed, not to claim a precision multi-parameter fit.

\subsection{Sampling and posterior results}
\label{subsec:revised_results}

Following the same prior-sensitivity strategy used in our timing analysis, posterior sampling is performed with an affine-invariant Goodman--Weare stretch-move ensemble sampler of the \texttt{emcee} type \cite{ForemanMackey2013,Huijser2017}. For each of the two choices of $\Pi_{\Gamma}$ we use 16 walkers and evolve every walker for $10^5$ steps. The first $2\times10^4$ steps are discarded as burn-in and every 25th post-burn-in state is retained. Each prior choice therefore yields 51,200 retained posterior samples.

For the primary uniform-prior run the mean acceptance fraction is 0.451. The integrated autocorrelation times of the five sampled parameters lie between about 106 and 128 steps, so the post-burn-in chain length is more than $6.2\times10^2$ autocorrelation times for every parameter. For the truncated-Gaussian sensitivity run the mean acceptance fraction is 0.478 and the autocorrelation times are approximately 94--108 steps, corresponding to more than $7.4\times10^2$ autocorrelation times. Trace plots were inspected for both calculations and show no persistent post-burn-in drift. These diagnostics demonstrate that the broad deformation posterior is not caused by insufficient chain length or poor sampling.

\begin{figure}[!htbp]
    \centering
    \includegraphics[width=0.95\linewidth]{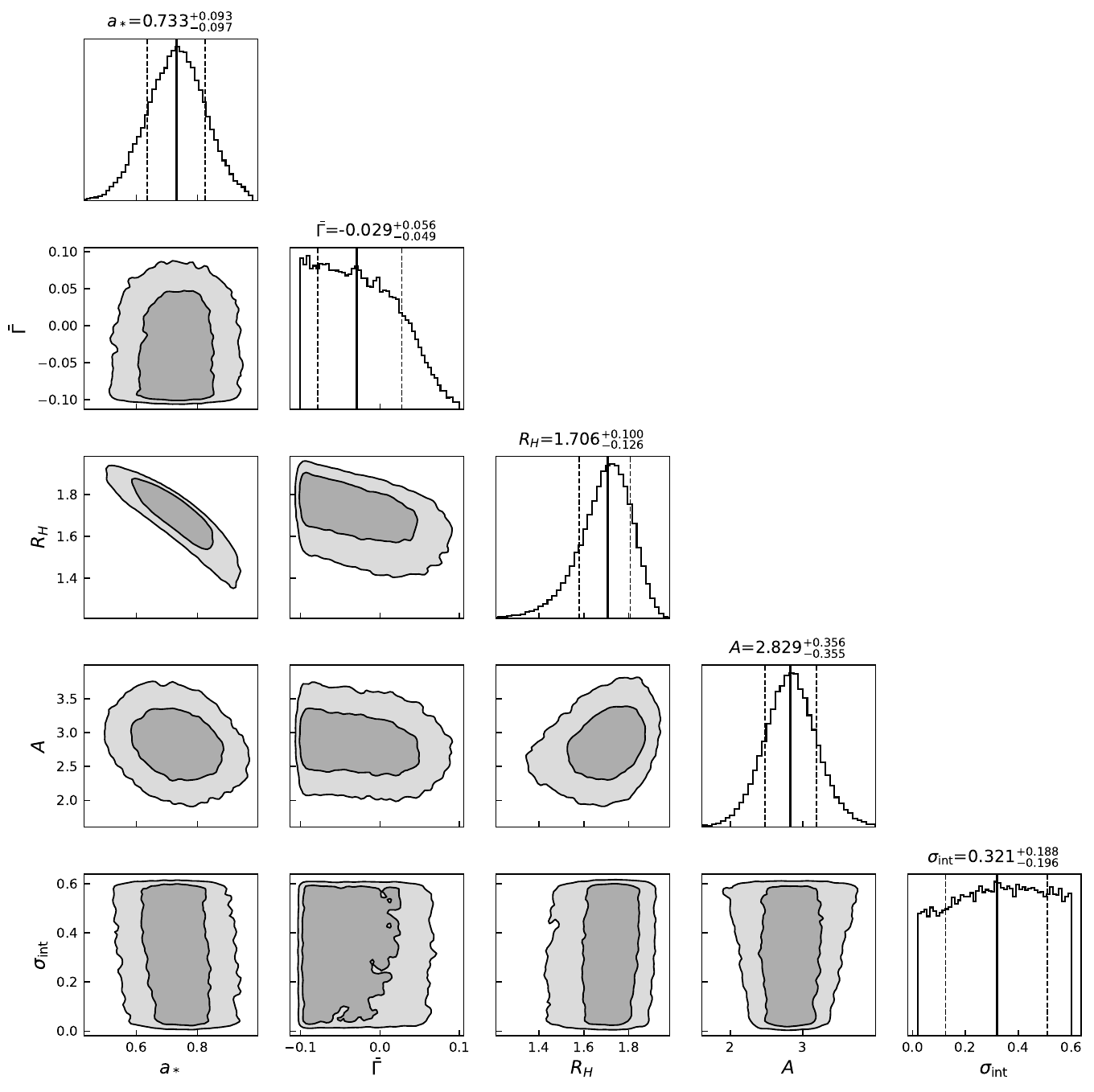}
    \caption{Posterior projection for GRO~J1655--40 under the fixed-flux baseline. The displayed quantities are $a_*$, the common effective hair amplitude $\bar{\Gamma}$, the derived horizon radius $R_H/M$, $A=\log_{10}K$, and $\sigma_{\rm int}$.}
    \label{fig:gro_corner}
\end{figure}

\begin{figure}[!htbp]
    \centering
    \includegraphics[width=0.95\linewidth]{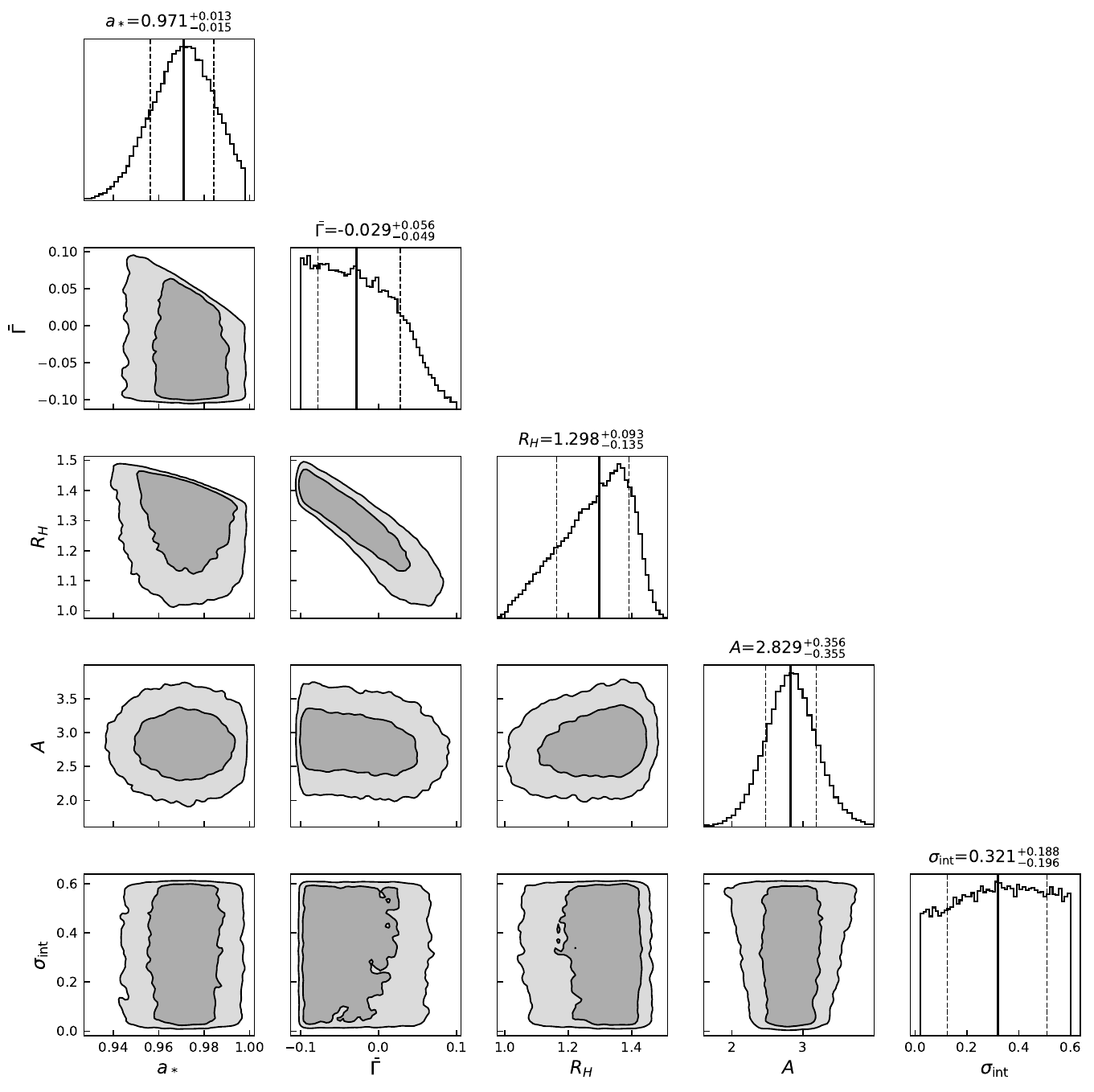}
    \caption{Posterior projection for GRS~1915+105 under the fixed-flux baseline. The high-spin proxy prior strongly intersects the horizon-existence boundary.}
    \label{fig:grs_corner}
\end{figure}

The source-wise posterior projections are shown in Figs.~\ref{fig:gro_corner} and \ref{fig:grs_corner}. The GRS~1915+105 posterior is additionally shaped by the intersection of its near-extremal spin prior with the horizon-existence boundary.

For GRO~J1655--40 the posterior remains well inside the horizon-admitting region over most of the adopted prior range. For GRS~1915+105, positive $\bar{\Gamma}$ values progressively remove the highest-spin part of the proxy prior. The upper side of the marginalized $\bar{\Gamma}$ posterior is therefore set mainly by the horizon-existence condition, not by a preference of the jet-power likelihood.

The marginalized fixed-flux posterior is summarized in Table~\ref{tab:revised_mcmc_results}.

\begin{table}[!htbp]
\centering
\caption{Posterior summary for the fixed-flux two-source Bayesian model. Horizon radii are derived quantities.}
\label{tab:revised_mcmc_results}
\begin{tabular}{lccc}
\hline
Parameter & Median & 16th percentile & 84th percentile \\
\hline
$\bar{\Gamma}$ & $-0.029$ & $-0.078$ & $0.027$ \\
$A=\log_{10}K$ & $2.829$ & $2.474$ & $3.185$ \\
$\sigma_{\rm int}$ & $0.321$ & $0.125$ & $0.509$ \\
$a_*^{\rm GRO}$ & $0.733$ & $0.636$ & $0.826$ \\
$R_H^{\rm GRO}/M$ & $1.706$ & $1.580$ & $1.807$ \\
$a_*^{\rm GRS}$ & $0.971$ & $0.956$ & $0.984$ \\
$R_H^{\rm GRS}/M$ & $1.298$ & $1.163$ & $1.391$ \\
\hline
\end{tabular}
\end{table}

\begin{figure}[!htbp]
    \centering
    \includegraphics[width=0.82\linewidth]{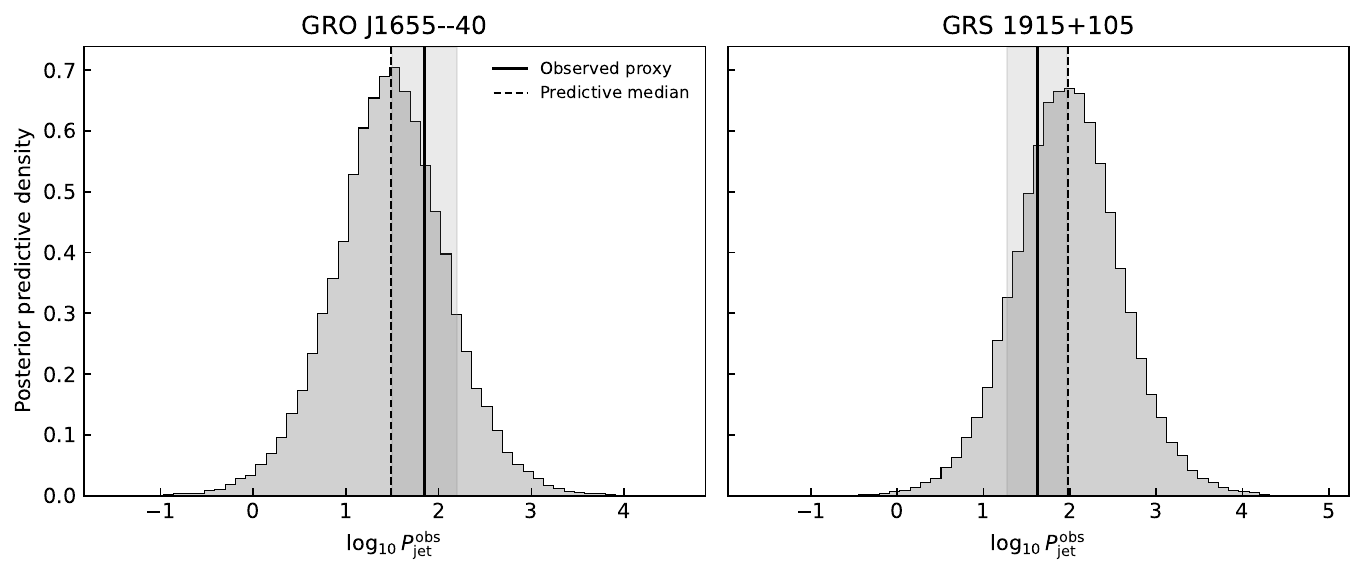}
    \caption{Posterior predictive distributions for the logarithmic observed jet-power proxies in the fixed-$\Phi_H$ baseline. For every retained posterior sample we draw a replicated datum from a Gaussian with variance $\sigma_{j}^{2}+\sigma_{\rm int}^{2}$, so these distributions include both the adopted observational uncertainty and the inferred intrinsic scatter. The solid vertical line marks the observed proxy, the light vertical band its quoted $1\sigma$ uncertainty, and the dashed line the predictive median.}
    \label{fig:two_source_predictive}
\end{figure}

\begin{figure}[!htbp]
    \centering
    \includegraphics[width=0.78\linewidth]{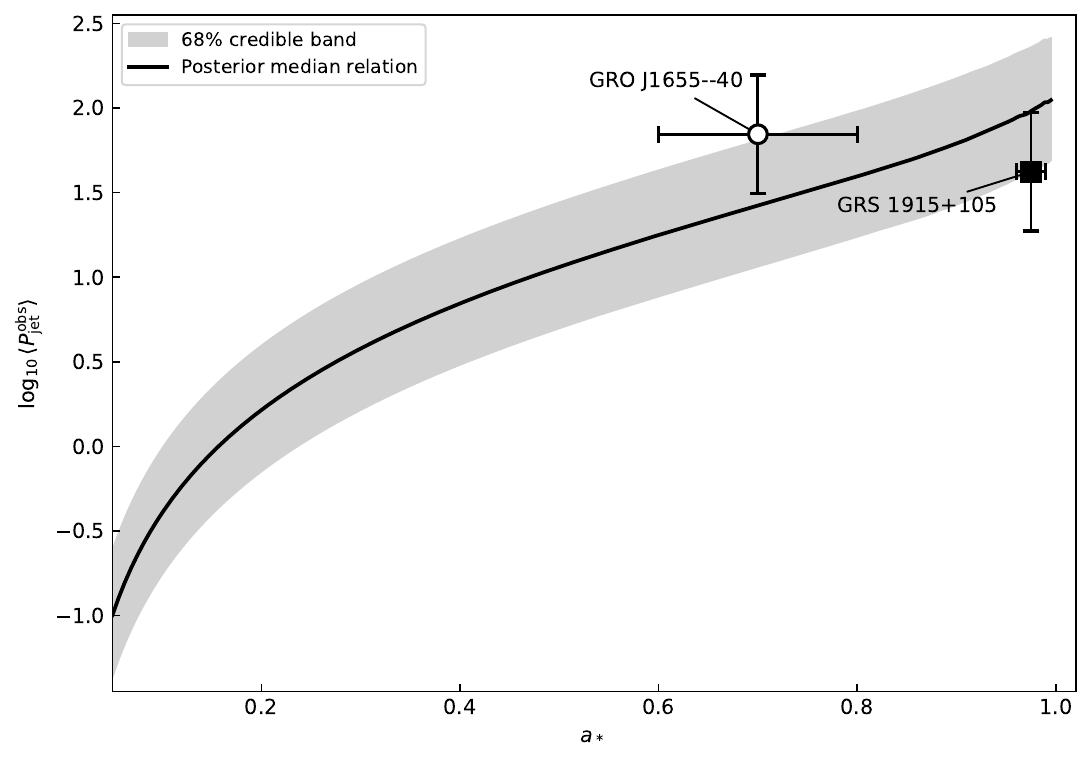}
    \caption{Latent fixed-$\Phi_H$ mean relation between the logarithmic jet-power proxy and spin. The solid curve is the posterior median and the shaded region is the 68\% credible band of the mean relation after marginalizing over $\bar{\Gamma}$ and the common normalization $A$. The directly labeled data points show the adopted measurements with $1\sigma$ uncertainties: GRO~J1655--40 is plotted as an open circle and GRS~1915+105 as a filled square. The band represents only the uncertainty of the mean relation; the additional predictive broadening from $\sigma_{\rm int}$ is shown separately in Fig.~\ref{fig:two_source_predictive}.}
    \label{fig:two_source_power_spin_fit}
\end{figure}

Thus the fixed-flux baseline gives
\begin{equation}
\bar{\Gamma}=-0.029^{+0.056}_{-0.049}.
\label{eq:revised_gamma_constraint}
\end{equation}
The posterior predictive distributions and the corresponding latent mean power--spin relation are shown in Figs.~\ref{fig:two_source_predictive} and \ref{fig:two_source_power_spin_fit}.

The replicated-data check is deliberately broader than the latent mean relation. The recomputed 16th--84th percentile posterior-predictive intervals are approximately $0.89$--$2.06$ for GRO~J1655--40 and $1.39$--$2.57$ for GRS~1915+105 in $\log_{10}P_{\rm jet}^{\rm obs}$. The observed values, $1.85$ and $1.62$, respectively, lie inside these intervals. Their posterior-predictive cumulative probabilities are about $0.74$ and $0.27$, so neither source is an obvious posterior-predictive outlier. At the same time, Fig.~\ref{fig:two_source_power_spin_fit} shows why this agreement should not be overinterpreted: a free normalization and substantial intrinsic scatter allow the two points to be accommodated without requiring a sharply determined $\bar{\Gamma}$.

\subsection{Prior sensitivity and identifiability of the deformation}
\label{subsec:prior_dominance}

A posterior interval by itself does not show whether the jet data identify $\bar{\Gamma}$, because the spin-localization factors and horizon-existence boundary already reshape the nominal deformation prior. We therefore construct, for each nominal prior in Eqs.~\eqref{eq:gamma_uniform_prior} and \eqref{eq:gamma_gaussian_prior}, the corresponding \emph{effective prior} by drawing the deformation and spin variables, applying the stated spin supports, and requiring a regular outer horizon, but omitting the jet-power likelihood.

For the primary uniform prior, the effective-prior and posterior 16th/50th/84th percentiles are
\begin{align}
\bar{\Gamma}_{\rm eff}^{\rm U}&=(-0.0756,-0.0235,0.0308),\\
\bar{\Gamma}_{\rm post}^{\rm U}&=(-0.0780,-0.0290,0.0273),
\end{align}
while the truncated-Gaussian sensitivity calculation gives
\begin{align}
\bar{\Gamma}_{\rm eff}^{\rm G}&=(-0.0626,-0.0193,0.0215),\\
\bar{\Gamma}_{\rm post}^{\rm G}&=(-0.0653,-0.0233,0.0180).
\end{align}
Thus the numerical location and width of the deformation posterior visibly follow the corresponding effective prior.

\begin{figure}[!htbp]
    \centering
    \includegraphics[width=0.95\linewidth]{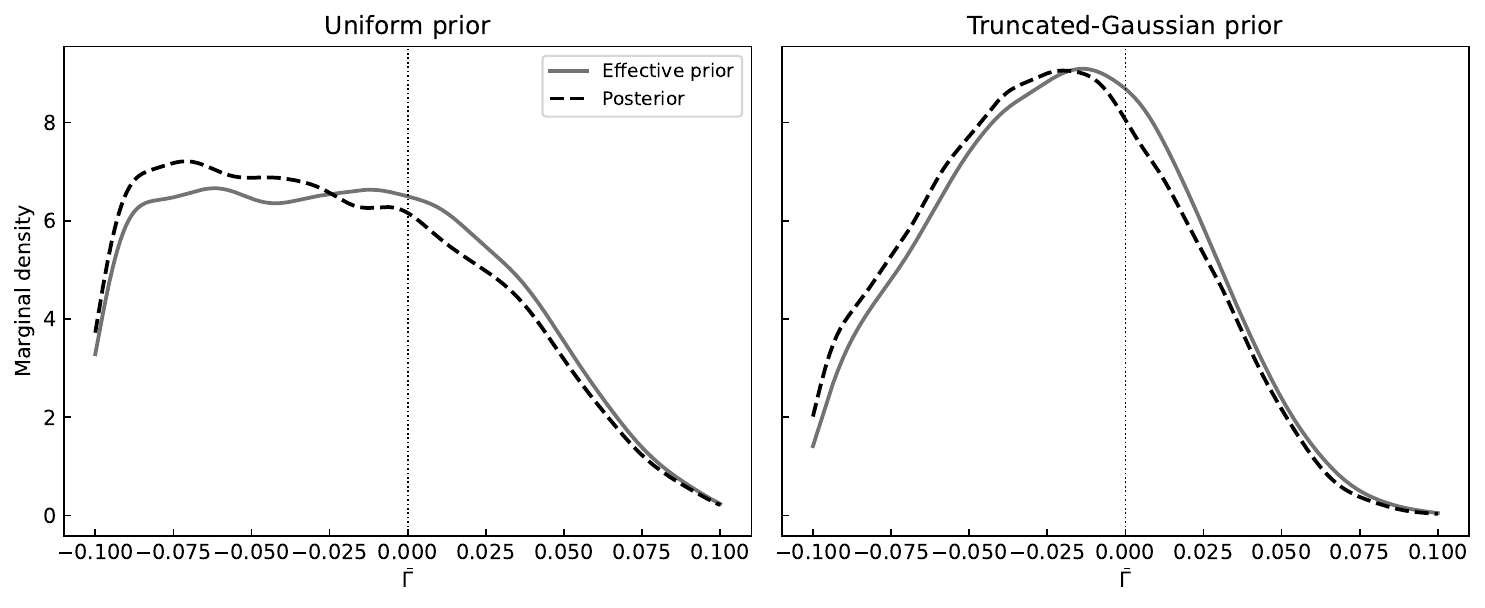}
    \caption{Controlled prior-sensitivity analysis for the common KR hair amplitude. The horizon-conditioned effective prior is compared with the corresponding marginal posterior for the primary uniform prior (left) and the truncated-Gaussian sensitivity prior (right). In both cases the posterior closely follows the effective prior, demonstrating negligible independent information from the two jet-power proxies.}
    \label{fig:gamma_prior_sensitivity}
\end{figure}

To quantify this comparison in the same way as the reference timing analysis, we define the central 68\% width ratio
\begin{equation}
R_{68}=\frac{\bar{\Gamma}^{\rm post}_{84}-\bar{\Gamma}^{\rm post}_{16}}
{\bar{\Gamma}^{\rm prior}_{84}-\bar{\Gamma}^{\rm prior}_{16}},
\label{eq:R68_gamma}
\end{equation}
and the normalized median shift
\begin{equation}
S_{\Gamma}=\frac{\bar{\Gamma}^{\rm post}_{50}-\bar{\Gamma}^{\rm prior}_{50}}
{\sigma_{\Gamma}^{\rm prior}}.
\label{eq:Sgamma}
\end{equation}
Here ``prior'' means the horizon-conditioned effective prior, since this is the distribution that reaches the likelihood after the physical-domain cut. The results are
\begin{table}[!htbp]
\centering
\caption{Controlled prior-sensitivity test for the common primary $s=3/2$ hair amplitude. $R_{68}\simeq1$ means negligible posterior contraction relative to the effective prior, while $|S_{\Gamma}|\ll1$ means that the jet likelihood does not appreciably displace the prior median.}
\label{tab:gamma_prior_sensitivity}
\begin{tabular}{lccc}
\hline
Nominal prior on $\bar{\Gamma}$ & Posterior $\bar{\Gamma}$ & $R_{68}$ & $S_{\Gamma}$ \\
\hline
Uniform & $-0.029^{+0.056}_{-0.049}$ & 0.990 & $-0.115$ \\
Truncated Gaussian & $-0.023^{+0.041}_{-0.042}$ & 0.990 & $-0.103$ \\
\hline
\end{tabular}
\end{table}
Both width ratios are within about one percent of unity and both median shifts are close to one tenth of an effective-prior standard deviation. Figure~\ref{fig:gamma_prior_sensitivity} displays this near coincidence directly.

As a prior-independent diagnostic we also calculate a jet-only profile chi-square. At fixed $\bar{\Gamma}$ we define
\begin{equation}
\chi^2_{\rm prof}(\bar{\Gamma})=
\min_{A,a_*^{\rm GRO},a_*^{\rm GRS}}
\sum_j\frac{[A+\log_{10}\widetilde{\mathcal P}_{\Phi,j}^{\rm KR}-\log_{10}P_{{\rm jet},j}^{\rm obs}]^2}{\sigma_{\log P,j}^2},
\label{eq:profile_chi_gamma}
\end{equation}
where the minimization uses the same top-hat spin supports and horizon domain, but no deformation or spin localization priors. We omit $\sigma_{\rm int}$ from this diagnostic so that an arbitrarily enlarged scatter cannot flatten the profile by construction. We then define $\Delta\chi^2_{\rm prof}=\chi^2_{\rm prof}-\min\chi^2_{\rm prof}$. Across the entire interval $-0.10\leq\bar{\Gamma}\leq0.10$ we find
\begin{equation}
\max\Delta\chi^2_{\rm prof}\simeq1.37\times10^{-2},
\label{eq:profile_chi_result}
\end{equation}
which is far below the usual $\Delta\chi^2\sim1$ scale associated with a one-parameter likelihood constraint. Representative values of the profile are summarized in Table~\ref{tab:gamma_profile_chi2}; the numerical minimum occurs at $\bar{\Gamma}\simeq0.049$, but the entire profile is so shallow that this location has no statistical significance.

\begin{table}[!htbp]
\centering
\caption{Representative values of the jet-only profile-likelihood diagnostic in the fixed-$\Phi_H$ model. The nuisance normalization and both source spins are profiled over the same physical supports used in the Bayesian analysis. The tiny values of $\Delta\chi^2_{\rm prof}$ throughout the full interval demonstrate the weak likelihood identifiability of $\bar{\Gamma}$.}
\label{tab:gamma_profile_chi2}
\begin{tabular}{ccc}
\hline
$\bar{\Gamma}$ & $\chi^2_{\rm prof}$ & $\Delta\chi^2_{\rm prof}$ \\
\hline
$-0.100$ & $1.3662\times10^{-2}$ & $1.3662\times10^{-2}$ \\
$-0.050$ & $5.465\times10^{-3}$ & $5.465\times10^{-3}$ \\
$0.000$ & $3.308\times10^{-7}$ & $3.308\times10^{-7}$ \\
$0.049$ & $1.82\times10^{-11}$ & $0$ \\
$0.050$ & $3.63\times10^{-10}$ & $3.45\times10^{-10}$ \\
$0.100$ & $2.68\times10^{-7}$ & $2.68\times10^{-7}$ \\
\hline
\end{tabular}
\end{table}

\subsection{Comparison with the literature-motivated $s=3$ benchmark}
\label{subsec:s3_crosscheck}

The original rotating-KR shadow analysis used $s=3$ as one representative non-Kerr case \cite{Kumar2020}. Because this member has a slower-than-$1/r$ deformation, we do not use it as the primary astrophysical benchmark. It nevertheless provides a useful check that the statistical diagnosis is not specific to the chosen radial exponent. Repeating the same fixed-$\Phi_H$ identifiability analysis with the $s=3$ member gives a uniform-prior posterior $\bar{\Gamma}_{3}=-0.027^{+0.057}_{-0.051}$, a width ratio $R_{68}=1.000$, and a nearly flat jet-only profile with $\max\Delta\chi^2_{\rm prof}\simeq1.34\times10^{-2}$.

\begin{table}[!htbp]
\centering
\caption{Robustness of the uniform-prior identifiability result to the radial exponent. The primary $s=3/2$ benchmark has a deformation that decays faster than the Kerr mass term; the $s=3$ row is retained only as a literature-motivated comparison. The dimensionless amplitudes have different mass scalings and are benchmark-specific quantities.}
\label{tab:s_benchmark_robustness}
\begin{tabular}{lcccc}
\hline
Benchmark & Asymptotic deformation & Posterior & $R_{68}$ & $\max\Delta\chi^2_{\rm prof}$ \\
\hline
$s=3/2$ (primary) & $r^{-4/3}$ & $-0.029^{+0.056}_{-0.049}$ & $0.990$ & $1.37\times10^{-2}$ \\
$s=3$ (comparison) & $r^{-2/3}$ & $-0.027^{+0.057}_{-0.051}$ & $1.000$ & $1.34\times10^{-2}$ \\
\hline
\end{tabular}
\end{table}

Both exponents give the same statistical diagnosis: the posterior is essentially prior-shaped and the jet-only profile is nearly flat. The weak jet-driven identifiability is therefore not an artifact of the asymptotic power chosen for the main benchmark. This comparison does not convert either posterior into a fundamental coupling bound because the deformation amplitudes carry different mass scalings and the rotating backgrounds remain model dependent.

The two diagnostics lead to the same conclusion. The medians obtained under the uniform and truncated-Gaussian priors must not be interpreted as two independent measurements of KR hair; they mainly reflect the corresponding effective priors after the high-spin horizon cut. The jet likelihood constrains correlated combinations of normalization, spin, scatter, and geometry, but it leaves $\bar{\Gamma}$ itself weakly identified.

The magnetic-flux prescription remains an additional physical systematic. The leading comparison in Sec.~\ref{subsec:flux_prescription_audit} shows that fixed total flux, a coordinate-radius proxy, and fixed local normal field correspond to physically different responses. Since the fixed-local-field factorization can remove the explicit horizon-radius dependence altogether, no deformation posterior obtained under one prescription should be promoted to a prescription-independent gravity constraint.

The two jet-power proxies thus do not provide a robust standalone bound on $\bar{\Gamma}$. Their main use here is as a controlled test of identifiability. A genuinely jet-driven KR constraint would require a larger source sample, source-specific magnetic-flux information or a physically motivated hierarchical flux model, self-consistent non-Kerr spin inference, and a magnetospheric calculation that determines how $\kappa_{\rm BZ}$ and the field geometry change in the KR spacetime.

\FloatBarrier
\section{Astrophysical implications}
\label{sec:astrophysical_implications}

The main astrophysical lesson is that the KR geometry alone does not determine a unique jet-power correction. The horizon angular velocity
\begin{equation}
\Omega_H^{\rm KR}=\frac{a}{r_H^2+a^2}
\end{equation}
is modified through the KR-dependent horizon radius, so a fixed total magnetic flux gives $P_{\rm BZ}\propto\Phi_H^2(\Omega_H^{\rm KR})^2$. However, the proper horizon area also changes. If the locally measured normal field is instead held fixed, the flux scales as $\Phi_H\simeq2\pi B_H(r_H^2+a^2)$ and the leading factorized dependence on $r_H$ cancels. Thus a claimed enhancement or suppression of BZ power is meaningful only after the magnetic quantity being compared has been specified.

This observation has direct consequences for source modeling. In realistic accretion systems the horizon flux is supplied and regulated by the accretion flow rather than imposed geometrically. In Kerr simulations, magnetic-flux saturation and magnetically arrested disk states play a central role in producing highly efficient jets \cite{Tchekhovskoy2011,McKinneyTchekhovskoyBlandford2012,WhiteStoneQuataert2019}, while reconnection, radiative effects, and magnetic topology modify the time-dependent outflow \cite{TchekhovskoyMcKinneyDexter2014,AvaraMcKinneyReynolds2016,Singh2018}. A comparable GRMHD or GRFFE calculation in the rotating KR spacetime is therefore needed to determine whether the global magnetosphere introduces residual KR dependence through $\kappa_{\rm BZ}$, flux saturation, field geometry, or plasma loading.

The Bayesian comparison also clarifies what can be inferred from the present microquasar jet proxies. Under both the uniform and truncated-Gaussian deformation priors, $R_{68}$ remains within about one percent of unity and $|S_{\Gamma}|\lesssim0.12$. Independently, the jet-only profile changes by only $\max\Delta\chi^2_{\rm prof}\simeq0.014$ across the full deformation interval. Together these diagnostics show that, in the present setup, the two jet-power proxies add essentially no independent information on $\bar{\Gamma}$. A more informative jet test must separate the spacetime prior, the magnetic-flux model, and the observational calibration of jet power.

Promising improvements include a larger source sample, source-specific estimates of magnetic flux, and hierarchical modeling in which the flux normalization is allowed to vary physically from source to source. Horizon-scale polarimetric information could be especially useful because it constrains the magnetic-field structure that is otherwise absorbed into the nuisance normalization. Combining such information with shadow and accretion-flow observables would provide a more discriminating strong-gravity test than jet power alone. The Event Horizon Telescope results for M87* and Sgr~A* motivate precisely this kind of multi-observable strategy \cite{EHT2019M87I,EHT2019M87V,EHT2022SgrAI}.

\FloatBarrier
\section{Conclusions}
\label{sec:conclusions}

We studied the leading Blandford--Znajek scaling in the power-law rotating Kalb--Ramond geometry of Ref.~\cite{Kumar2020}. The analysis was aimed at three issues that enter before a jet-power posterior can be read as a modified-gravity constraint: metric-level identifiability, the choice of magnetic-flux prescription, and the amount of information supplied by the jet data themselves. The rotating spacetime has been treated throughout as an adopted stationary background, not as a newly derived exact solution of the complete Einstein--KR system.

The clearest structural result already appears at the metric level. The $s=2$ member is exactly reducible to Kerr by redefining the mass parameter, so its hair amplitude is not an independent non-Kerr observable. In addition, for positive $s$ the power-law correction falls as $r^{-2/s}$; values $s>2$ therefore decay more slowly than the standard mass term. For this reason we use $s=3/2$ as the main benchmark; its $r^{-4/3}$ correction leaves the $1/r$ mass term asymptotically leading. Repeating the statistical test for the literature-used $s=3$ case gives the same prior-dominated result, so the weak identifiability is robust to this change of radial exponent.

The magnetic comparison shows that the leading BZ response is not determined by the geometry alone. With fixed total horizon flux, the deformation enters through the modified horizon angular velocity. If instead a local normal field is held fixed and the proper horizon area is used, the leading explicit horizon-radius dependence cancels. A coordinate-radius flux proxy produces yet another trend. Apparent enhancement or suppression of the jet power is therefore meaningful only after the magnetic quantity being held fixed has been specified. A global KR-specific force-free or GRMHD solution would be required to determine whether field geometry, flux saturation, plasma loading, or the BZ coefficient introduces additional deformation dependence.

The statistical analysis points in the same direction. For both the primary uniform deformation prior and the controlled truncated-Gaussian alternative, the marginal posterior closely follows the corresponding horizon-conditioned effective prior. The central 68\% width ratios are $0.990$ in both runs, while the normalized median shifts are only $-0.115$ and $-0.103$, respectively. The independent jet-only profile likelihood is also nearly flat across the full deformation interval, with a maximum variation of only about $1.4\times10^{-2}$ in profile chi-square. The resulting posterior intervals should therefore be read as compatibility regions shaped mainly by the adopted prior structure, the high-spin horizon boundary, the common normalization, and the intrinsic scatter, rather than as independent jet-driven measurements of the KR hair amplitude.

Taken together, these results make the present work an identifiability and systematics study rather than a standalone observational bound on Kalb--Ramond gravity. The analysis isolates the degenerate $s=2$ slice, shows explicitly how the leading jet response depends on the magnetic prescription, and separates posterior localization from likelihood information. The same weak-identifiability pattern in the $s=3/2$ and $s=3$ cases indicates that this conclusion is not tied to one chosen exponent. A quantitatively stronger jet test will need a rotating KR background with controlled asymptotics and conserved charges, self-consistent non-Kerr source inference, source-specific or hierarchical magnetic-flux modeling, more systems, and ultimately a global magnetospheric calculation in the same spacetime.

\section*{Acknowledgments}
S.M. gratefully acknowledges support from Grant FZ-20200929385 of the Agency of Innovative Developments of the Republic of Uzbekistan.

\section*{Data Availability}
No new observational data were generated in this study. The observational inputs used in the two-source consistency analysis are taken from the works cited in Sec.~\ref{sec:MCMC_constraints}. The custom numerical scripts and derived numerical data supporting the figures, magnetic-flux comparisons, and Bayesian calculations are being prepared for public archival deposit and are available from the corresponding author upon reasonable request in the meantime.

\bibliography{reference}

\end{document}